\documentclass[sigconf]{acmart}
\usepackage{tikz}% for circle box

\usepackage{booktabs}
\usepackage{array}
\usepackage{balance} 
\usepackage{lipsum}
\usepackage{multirow}
\usepackage{booktabs}
\usepackage{amsmath}
\newcommand{\ie}{\emph{i.e., }}
\newcommand{\eg}{\emph{e.g., }}
\newcommand{\etal}{\emph{et al. }}

\newcommand{\wrt}{\emph{w.r.t. }}
\newcommand{\cf}{\emph{cf. }}

\usepackage{subfigure}

\usepackage{algorithm}  
\usepackage{algorithmicx}  
\usepackage{algpseudocode}  
\usepackage{amsmath}  
\usepackage{enumitem}
\usepackage{tabularx}

\usepackage[utf8]{inputenc}
\usepackage[english]{babel}

\usepackage{amsthm}

\usepackage{bm}

\newtheorem{theorem}{Theorem}[section]

\begin{document}

% \copyrightyear{}
% \acmYear{2021} 
% \acmConference[]{}{}{}
% \acmBooktitle{}
% \acmPrice{15.00}

\copyrightyear{2021}
\acmYear{2021}
\setcopyright{acmcopyright}\acmConference[KDD '21]{Proceedings of the 27th ACM SIGKDD Conference on Knowledge Discovery and Data Mining}{August 14--18, 2021}{Virtual Event, Singapore}
\acmBooktitle{Proceedings of the 27th ACM SIGKDD Conference on Knowledge Discovery and Data Mining (KDD '21), August 14--18, 2021, Virtual Event, Singapore}
\acmPrice{15.00}
\acmDOI{10.1145/3447548.3467249}
\acmISBN{978-1-4503-8332-5/21/08}

\fancyhead{}
% \title{Deconfounded Recommender System for \\Alleviating Bias Amplification}
\title{Deconfounded Recommendation for Alleviating\\ Bias Amplification}

% \subtitle{Submission ID: 1441}
% \author{Anonymous authors}

\author{Wenjie Wang$^{1}$, Fuli Feng$^{12*}$, Xiangnan He$^{3}$, Xiang Wang$^{12}$, and Tat-Seng Chua$^1$}

\def\authors{Wenjie Wang, Fuli Feng, Xiangnan He, Xiang Wang, and Tat-Seng Chua}

\affiliation{
\institution{$^1$National University of Singapore, $^2$Sea-NExT Joint Lab, $^3$University of Science and Technology of China}
\country{}
}
\email{{wenjiewang96, fulifeng93,xiangnanhe}@gmail.com, xiangwang@u.nus.edu, dcscts@nus.edu.sg}

\thanks{$*$ Corresponding author: Fuli Feng (fulifeng93@gmail.com). This research is supported by the Sea-NExT Joint Lab, the National Natural Science Foundation of China (61972372), and National Key Research and Development Program of China (2020AAA0106000).}

% \author{Wenjie Wang}
% \email{wenjiewang96@gmail.com}
% \affiliation{%
%   \institution{National University of Singapore}
% }
% \author{Fuli Feng}
% \email{fulifeng93@gmail.com}
% \affiliation{%
%   \institution{National University of Singapore}
% }

% \author{Xiangnan He}
% \email{xiangnanhe@gmail.com}
% \affiliation{%
%   \institution{University of Science and Technology of China}
% }
% \author{Hanwang Zhang}
% \email{hanwangzhang@ntu.edu.sg}
% \affiliation{%
%   \institution{Nanyang Technological University}
% }

% \author{Tat-Seng Chua}
% \email{dcscts@nus.edu.sg}
% \affiliation{%
%   \institution{National University of Singapore}
% }

\begin{abstract}

Recommender systems usually amplify the biases in the data. The model learned from historical interactions with imbalanced item distribution will amplify the imbalance by over-recommending items from the major groups. Addressing this issue is essential for a healthy ecosystem of recommendation in the long run. Existing works apply bias control to the ranking targets (\eg calibration, fairness, and diversity), but ignore the true reason for bias amplification and trade off the recommendation accuracy.

In this work, we scrutinize the cause-effect factors for bias amplification, identifying the main reason lies in the \textit{confounder} effect of imbalanced item distribution on user representation and prediction score. The existence of such confounder pushes us to go beyond merely modeling the conditional probability and embrace the causal modeling for recommendation. Towards this end, we propose a \textit{Deconfounded Recommender System} (DecRS), which models the causal effect of user representation on the prediction score. The key to eliminating the impact of the confounder lies in \textit{backdoor adjustment}, which is however difficult to do due to the infinite sample space of the confounder. For this challenge, we contribute an approximation operator for backdoor adjustment which can be easily plugged into most recommender models. Lastly, we devise an inference strategy to dynamically regulate backdoor adjustment according to user status. We instantiate DecRS on two representative models FM~\cite{rendle2010factorization} and NFM~\cite{he2017nfm}, and conduct extensive experiments over two benchmarks to validate the superiority of our proposed DecRS.

\end{abstract}

\begin{CCSXML}
<ccs2012>
   <concept>
       <concept_id>10002951.10003317.10003347.10003350</concept_id>
       <concept_desc>Information systems~Recommender systems</concept_desc>
       <concept_significance>500</concept_significance>
       </concept>
   <concept>
       <concept_id>10002951.10003260.10003261.10003269</concept_id>
       <concept_desc>Information systems~Collaborative filtering</concept_desc>
       <concept_significance>500</concept_significance>
       </concept>
 </ccs2012>
\end{CCSXML}

\ccsdesc[500]{Information systems~Recommender systems}
\ccsdesc[500]{Information systems~Collaborative filtering}

\keywords{Deconfounded Recommendation, User Interest Imbalance, Bias Amplification}

\maketitle

\vspace{-0.3cm}
\section{Introduction}

\begin{figure}[tb]
\setlength{\abovecaptionskip}{0.1cm}
\setlength{\belowcaptionskip}{-0.5cm}
% \vspace{}
\centering
\includegraphics[scale=0.56]{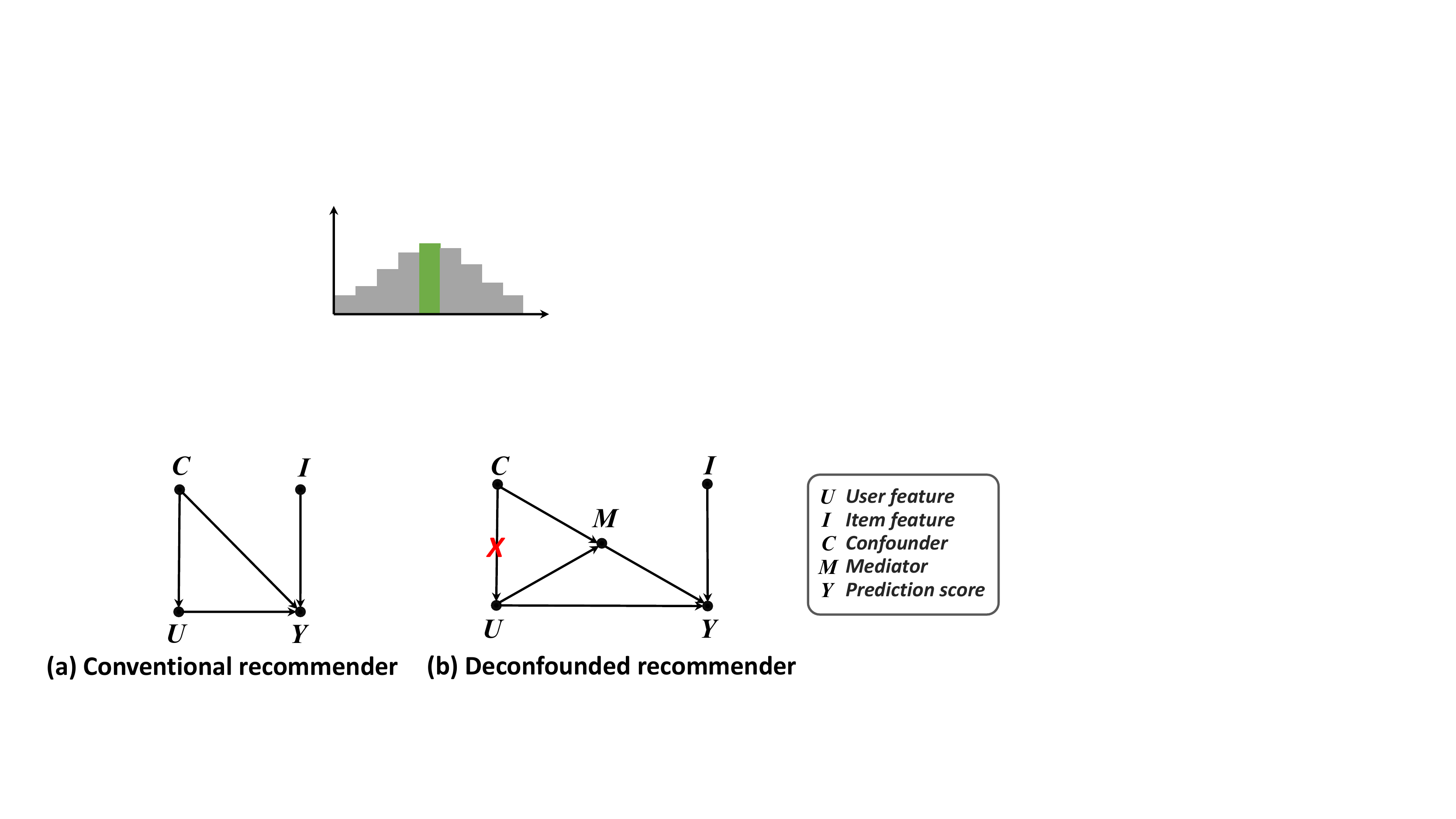}
\caption{Illustration of bias amplification.}
\label{fig:example}
\end{figure}

Recommender System (RS) has been widely used to achieve personalized recommendation in most online services, such as social networks and advertising~\cite{wang2019NGCF}.
% CF performs the personalized recommendation by the assumption that similar users would prefer similar items.
% A basic assumption of CF is that similar users would prefer similar items.
%Generally, CF models are collecting users' historical interactions (\eg clicks and purchase) to learn the user interest.
Its default choice is to learn user interest from historical interactions (\eg clicks and purchases), which typically exhibit data bias, \ie the distribution over item groups (\eg the genre of movies) is imbalanced.
%However, a serious issue in existing CF models is \textit{bias amplification}~\cite{steck2018calibrated}: users' interactions over item groups (\eg the genre of movies) are imbalanced in the history, and the recommender models tend to amplify the imbalance in the new recommendation list. 
Consequently, recommender models face the \textit{bias amplification} issue~\cite{steck2018calibrated}: over-recommending the majority group and amplifying the imbalance. 
Figure \ref{fig:example}(a) illustrates this issue with an example in movie recommendation, where 70\% of the movies watched by a user are action movies, but action movies take 90\% of the recommendation slots.
%For instance, as shown in Figure \ref{fig:example}(a), one user watched 70\% action movies and 30\% romance movies in the history, and then the model would recommend more action movies (\eg 90\%)~\cite{steck2018calibrated}. 
Undoubtedly, over-emphasizing the items from the majority groups will limit a user's view and decrease the effectiveness of recommendations. 
%What’s even worse, such bias amplification will be increasingly serious due to the feedback loop~\cite{chaney2018algorithmic}, which will decrease the recommendation accuracy and induce the unfairness and low diversity, possibly leading to the issues of filter bubbles~\cite{nguyen2014exploring} and echo chambers~\cite{Ge2020Understanding}.
Worse still, due to feedback loop~\cite{chaney2018algorithmic}, such bias amplification will intensify with time, causing more issues like filter bubbles~\cite{nguyen2014exploring} and echo chambers~\cite{Ge2020Understanding}.
Existing works alleviate bias amplification by introducing bias control into the ranking objective of recommender models, which are mainly from three perspectives:
%1) fairness in recommendation pursues the similar treatments (\eg exposure or click) for different user/item groups~\cite{singh2018fairness}. Preserving the exposure fairness across different item groups might reduce bias amplification~\cite{Morik2020Controlling};
1) fairness~\cite{singh2018fairness,Morik2020Controlling}, which pursues equal exposure opportunities for items of different groups;
% fairness across item groups regarding the exposure opportunities, and undoubtedly alleviates the bias amplication issue. 
2) diversity~\cite{chandar2013preference}, which intentionally increases the covered groups in a recommendation list,
% penalizes the similar recommendations for the user, including the items from the same group, is thus able to reduce the recommendations of the items in the major group.
%diversity of recommendations focuses on recommending dissimilar items to each user, where the similarity could be measured by the item group (\eg the movie genre). These works may reduce the recommendations of the items in the majority group, \eg action movies in Figure \ref{fig:example}(a);
%3) a more relevant research direction towards bias amplification is called \textit{calibrated recommendation}~\cite{steck2018calibrated}. The goal is to recommend different groups of items in the same proportion with that of the historical interactions by re-ranking (\eg 7:3 in Figure \ref{fig:example}(a)). 
and 3) calibration~\cite{steck2018calibrated}, which encourages the distribution of recommended item groups to follow that of interacted items of the user. 
% encourages the recommendations for a user to follow the same distribution over the item groups as the user's historical interactions. 
% However, although these works, especially calibrated recommendation, can alleviate bias amplification, they inevitably have the price of sacrificing accuracy~\cite{steck2018calibrated, singh2018fairness}. 
% This might because most works totally ignore bias amplification comes from the training process, leaving the key theoretical question unanswered: why does the recommender training amplify the imbalance in the history?
However, these methods alleviate bias amplification at the cost of sacrificing recommendation accuracy~\cite{steck2018calibrated, singh2018fairness}. 
%%%
% Not coherent with the previous sentence
%%%
%This might because most works totally ignore bias amplification comes from the conventional CF framework, leaving the key theoretical question unanswered: what is the root reason of bias amplification?
More importantly, the fundamental question is not answered: what is the root reason for bias amplification?

% To this end, we propose a causal graph to analyze the causal relations in CF models, 
%To this end, we inspect the cause-effect factors in CF models, and find that the user history distribution over item groups (\eg $[0.7, 0.3]$ in Figure \ref{fig:example}(a)) is a \textit{confounder}~\cite{pearl2009causality}, which results in bias amplification. 
After inspecting the cause-effect factors in recommender modeling, we attribute bias amplification to a \textit{confounder}~\cite{pearl2009causality}. The historical distribution of a user over item groups (\eg $[0.7, 0.3]$ in Figure \ref{fig:example}(a)) is a confounder between the user's representation and the prediction score.
In the conventional RS, the user/item features (\eg ID and attributes) are first embedded into the representation vectors, which are then fed into an interaction module (\eg factorization machines (FM)~\cite{rendle2010factorization}) to calculate the prediction score for the user-item pair~\cite{He2017Neural}. In other words, recommender models estimate the \textit{conditional probability} of clicks given user/item representations. 
% During training, the parameters in user/item representations and the interaction module are optimized to reconstruct historical interactions. 
From a causal view, user and item representations can be regarded as the causes of the prediction score, and the interaction module should encode the causal relations between them~\cite{pearl2009causality}. 
% By inspecting the causal relations, we find that a hidden variable is implicitly learned from the imbalanced historical data during training:
% By inspecting the causal relations, we find that a hidden confounder implicitly affects both the user representation and the prediction score during training: the user history distribution over item groups. 
But inspecting the causal relations, we find that the hidden confounder, \ie the user historical distribution over item groups, affects both the user representation and the prediction score.
Due to the modeling of conditional probability, recommender models are affected by the confounder and thus suffer from a spurious correlation between the user and the prediction score. That is, 
given two item groups, the one that the user interacted more in the history will receive higher prediction scores, even though their items have the same matching level. 
% given the item in a group (\eg action movies), the more items in this group that the user has interacted with in the history, the higher the prediction score is likely to be. 
%The empirical results of FM on ML-1M dataset\footnote{\url{https://grouplens.org/datasets/movielens/1m/.}} in Figure \ref{fig:example}(b) verify the correlation: among the items with the same ratings (\eg ratings = 4), the ones in the majority group receive higher prediction scores. Therefore, the items in the majority group, even including undesirable or low-quality ones (see example in Figure \ref{fig:example}(c)), could occupy the recommendation opportunities of the items in the minority group.
Figure \ref{fig:example}(b) shows empirical evidence from the FM on ML-1M dataset:
% \footnote{\url{https://grouplens.org/datasets/movielens/1m/.}}
among the items with the same ratings (\eg ratings = 4), the ones in the majority group will receive higher prediction scores. Therefore, the items in the majority group, even including those undesirable or low-quality ones (see example in Figure \ref{fig:example}(c)), could deprive the recommendation opportunities of the items in the minority group.

% we find that a hi dden variable is implicitly learned from the imbalanced historical data during training: the user history distribution over item groups (\eg $[0.7, 0.3]$ in Figure \ref{fig:example}(a)), which actually acts as a \textit{confounder}~\cite{pearl2009causality} and results in bias amplification during recommender training. 

% Indeed, this correlation aligns with the core idea of CF models, recommending items similar to the historical ones. It helps to exclude the item groups that users dislike, \eg horror movies in the example of Figure \ref{fig:example}(a). 
% However, the correlation also induces bias amplification: some items with lower ratings in the majority group might have a larger prediction score than the ones with higher ratings (see the example in Figure \ref{fig:example}(c)). Therefore, the items in the majority group, even including undesirable or low-quality ones, occupy the recommendation opportunities of the items in the minority group.

% We find that the user history distribution over categories is a confounder, which affects both the 

% The key of addressing bias amplification lies in eliminating the spurious correlation during the recommender model training. To achieve this goal, we propose a model-agnostic deconfounded collaborative filtering (DeCF) framework, which empowers existing CF models with the ability of causal inference to alleviate bias amplification.
% In particular, 1) during the training process, we explicitly model the causal relations in the recommender model. In addition to conventional training, DeCF applies deconfounded training with backdoor adjustment~\cite{pearl2009causality} to remove the effect of confounder on user representation learning, which significantly alleviates bias amplification.
% 2) In the phase of inference, considering the correlation helps to exclude the irrelevant item groups for some users, DeCF leverages an auto-adjusting inference strategy to combine conventional training and deconfounded training for personalized item ranking.
% Finally, we instantiate DeCF on two representative CF models and extensive experiments over two benchmarks demonstrate that the proposed DeCF framework significantly improves the recommendation accuracy of existing CF models while alleviating bias amplification. 
% The code and data are released to facilitate the research in this field\footnote{\url{https://anonymous.4open.science/r/DeCF/.}}. 

% The key of addressing bias amplification lies in eliminating the spurious correlation during the recommender model training. 
The key to addressing bias amplification lies in eliminating the spurious correlation in the recommender modeling. 
To achieve this goal, we need to push the conventional RS to go beyond modeling the conditional probability and embrace the causal modeling of user representation on the prediction score. We propose a novel \textit{Deconfounded Recommender System} (DecRS), which
% which incorporates backdoor adjustment to eliminate the impact of the confounder. 
explicitly models the causal relations during training, and leverages backdoor adjustment~\cite{pearl2009causality} to eliminate the impact of the confounder. However, the sample space of the confounder is huge, making the traditional implementation of backdoor adjustment infeasible. To this end, we derive an approximation of backdoor adjustment, which is universally applicable to most recommender models. 
% 2) In the phase of inference, considering the correlation helps to exclude the irrelevant item groups for some users, 
Lastly, we propose a user-specific inference strategy to dynamically regulate the influence of backdoor adjustment based on the user status.
We instantiate DecRS on two representative models FM~\cite{rendle2010factorization} and neural factorization machines (NFM)~\cite{he2017nfm}.
Extensive experiments over two benchmarks demonstrate that 
% the proposed DeCF framework significantly improves the recommendation accuracy of existing CF models while alleviating bias amplification. 
our DecRS not only alleviates bias amplification effectively, but also improves the recommendation accuracy over the backbone models. 
% The code and data are released to facilitate the research in this emerging field\footnote{\url{https://anonymous.4open.science/r/6c28d9a8-10c3-40d3-99f1-66c4684f85d8/.}}. 

Overall, the main contributions of this work are threefold:
\begin{itemize}[leftmargin=*]
    \item We construct a causal graph to analyze the causal relations in recommender models, which reveals the cause of bias amplification from a causal view.
    % \item We propose a novel DeCF framework with a new deconfounded training method and an auto-adjusting inference strategy, which removes the effect of the confounder dynamically.
    \item We propose a novel DecRS with an approximation of backdoor adjustment to eliminate the impact of the confounder, which can be incorporated into existing recommender models to alleviate bias amplification.
    \item We instantiate DecRS on two representative recommender models, and conduct extensive experiments on two benchmarks which validate the effectiveness of our proposal.
    
\end{itemize}

\section{Methodology}
\label{sec:method}
%In this section, we will first analyze the conventional RS and explain the reason of bias amplification from a causal view. It is followed by the introduction of the proposed DecRS. 
In this section, we first analyze the conventional RS from a causal view and explain the reason for bias amplification, which is followed by the introduction of the proposed DecRS.

\subsection{A Causal View on Bias Amplification}
To study bias amplification, we build up a causal graph to explicitly analyze the causal relations in the conventional RS.

\begin{figure}[tb]
\setlength{\abovecaptionskip}{0cm}
\setlength{\belowcaptionskip}{-0.5cm}
% \vspace{}
\centering
\includegraphics[scale=0.55]{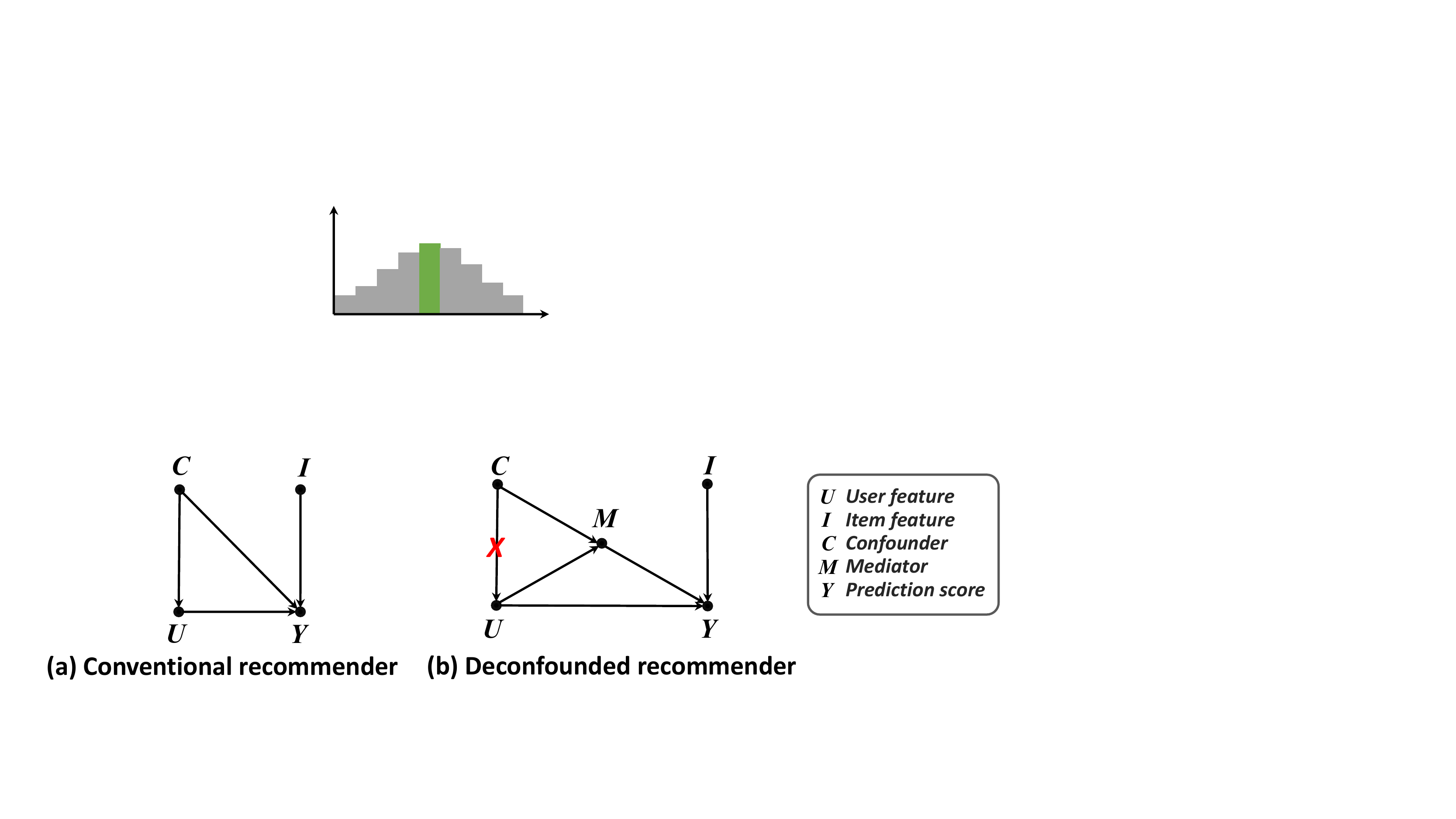}
\caption{(a) The causal graph of conventional RS. (b) The causal graph used in DecRS.}
\label{fig:causal_graph}
\end{figure}

% \vspace{-0.15cm}
% \paragraph{\textbf{Causal Graph}}
\subsubsection{Causal Graph}

% In the CF modeling, user and item features (\eg ID, gender, and movie genre) are first embedded into the vectors as the user and item representations (\ie $U$ and $I$), respectively. Then CF models estimate the prediction score $Y$ by the conditional probability $P(Y|U, I)$~\cite{rendle2010factorization, he2017nfm}. However, a hidden variable, \ie the user's historical distribution over item groups $D$, will implicitly affect the user representation $U$ and the prediction score $Y$ when the models are trained on the imbalanced data. 

%Formally, we construct a causal graph to scrutinize the causal relations in recommender models. 
% Figure~\ref{fig:causal_graph}(a) shows the causal graph 
We scrutinize the causal relations in recommender models and abstract a causal graph, as shown in Figure \ref{fig:causal_graph}(a), which consists of five variables: $U$, $I$, $D$, $M$, and $Y$. Note that we use the capital letter (\eg $U$), lowercase letter (\eg $\bm{u}$), and letter in the calligraphic font (\eg $\mathcal{U}$) to represent a variable, its particular value, and its sample space, respectively. In particular,
\begin{itemize}[leftmargin=*]
    \item $U$ denotes user representation. For one user, $\bm{u} = [\bm{u}_1, ..., \bm{u}_K]$ represents the embeddings of $K$ user features (\eg ID, gender, and age)~\cite{rendle2010factorization}, where $\bm{u}_k \in \mathbb{R}^H$ is one feature embedding. 
    \item $I$ is item representation and each $\bm{i}$ denotes the embeddings of several item features (\eg ID and genre) which are similar to $\bm{u}$.
    \item $D$ represents the user historical distribution over item groups. Groups can be decided by item attributes or similarity~\cite{steck2018calibrated}. Given $N$ item groups $\{g_1, ..., g_N\}$, $\bm{d}_u = [p_u(g_1), ..., p_u(g_N)] \in \mathbb{R}^N$ is a particular value of $D$ when the user is $u$, where $p_u(g_n)$ is the click frequency of user $u$ over group $g_n$ in the history\footnote{In this work, we use click to represent any implicit feedback, such as purchase and watch. For brevity, $u$ and $i$ may be used to denote the user and item, respectively. Besides, $n$ is used to represent any value in $\{1, 2,..., N\}$.}. For instance, for the user $u$ in Figure \ref{fig:example}(a), $\bm{d}_u$ is $[0.7, 0.3]$ if $N=2$.
    % \item $M$ describes how much users like each item group in the history, represented by $N$ vectors $\bm{m}_u=[\bm{m}_{u, 1}, ..., \bm{m}_{u, N}]$, where $\bm{m}_{u, n}$ represents the user $u$'s historical interest in group $g_n$\footnote{$n$ can be used to represent any value in $\{1, 2,..., N\}$.}.
    \item $M$ is the group-level user representation. A particular value $\bm{m}\in \mathbb{R}^H$ is a vector which describes how much the user likes different item groups. $\bm{m}$ can be obtained from the values of $U$ and $D$. That is, $M$ is deterministic if $U$ and $D$ are given so that we can represent $\bm{m}$ by a function $M(\bm{d},\bm{u})$ with $\bm{d}$ and $\bm{u}$ as inputs. 
    To keep generality, we incorporate $M$ into the causal graph because many recommender models (\eg FM) have modeled the user preference over item groups explicitly or implicitly by using the group-related features (\eg movie genre). 
    \item $Y$ with $y\in [0,1]$ is the prediction score for the user-item pair.
\end{itemize}

The edges in the graph describe the causal relations between variables, \eg $U \rightarrow Y$ means that $U$ has a direct \textit{causal effect}~\cite{pearl2009causality} on $Y$, \ie changes on $U$ will affect the value of $Y$. In particular,
\begin{itemize}[leftmargin=*]
    \item $D \rightarrow U$: the user historical distribution over item groups affects user representation $U$, making it favor the group with a higher click frequency (\ie majority group). This is because user representation is optimized to fit the imbalanced historical data.
    
    \item $(D, U) \rightarrow M$: $D$ and $U$ decide the group-level user representation. 
    
    \item $(U, M, I) \rightarrow Y$: The edges show that $U$ affects $Y$ by two paths: 1) the direct path $U \rightarrow Y$ which denotes the user's pure preference over the item, and 2) the indirect path $U \rightarrow M \rightarrow Y$, indicating that the prediction score could be high because the user shows interest in the item group rather than the item.
\end{itemize}
According to the causal theory~\cite{pearl2009causality}, since $D$ affects both $U$ and $Y$, $D$ is a \textit{confounder} between $U$ and $Y$, resulting in the spurious correlation when estimating the correlation between $U$ and $Y$. 
%Besides, $M$ is a \textit{mediator} between $U$ and $Y$, which means that $U$ affects $Y$ via the mediator $M$.

% \vspace{-0.15cm}
% \paragraph{\textbf{Conventional Training}} 
\subsubsection{Conventional RS}
% Existing recommender models take user and item representations as inputs to estimate the prediction score \ie $P(Y|U, I)$.
% which are trained on the users' historical data with the imbalanced distribution $D$. As such, they suffer from the spurious correlation caused by the confounder $D$.
% However, user representation $U$ and the prediction score $Y$ are implicitly affected by the confounder $D$ because the recommender models are trained on the imbalanced historical data over item groups. 
Due to the confounder, existing recommender models that estimate the conditional probability $P(Y|U, I)$ face the spurious correlation, which leads to bias amplification. 
Formally, given $U=\bm{u}$ and $I=\bm{i}$, we can derive the conditional probability $P(Y|U, I)$ by:
\begin{subequations}
\label{equ:P_Y_UI}
\begin{align}\footnotesize
& P(Y|U=\bm{u}, I=\bm{i}) \notag \\
            &= \frac{\sum_{\bm{d}\in \mathcal{D}} \sum_{\bm{m}\in \mathcal{M}} P(\bm{d})P(\bm{u}|\bm{d})P(\bm{m}|\bm{d}, \bm{u})P(\bm{i})P(Y|\bm{u}, \bm{i}, \bm{m})}{P(\bm{u})P(\bm{i})} \\
            &= \sum_{\bm{d}\in \mathcal{D}}\sum_{\bm{m}\in \mathcal{M}} P(\bm{d}|\bm{u})P(\bm{m}|\bm{d},\bm{u})P(Y|\bm{u},\bm{i},\bm{m}) \\
            &= \sum_{\bm{d}\in \mathcal{D}} P(\bm{d}|\bm{u})P(Y|\bm{u},\bm{i},M(\bm{d},\bm{u})) \\
            &= P(\bm{d}_u|\bm{u})P(Y|\bm{u},\bm{i},M(\bm{d}_u,\bm{u})),
\end{align}
\end{subequations}
%where $\mathcal{D}$ and $\mathcal{M}$ are the sets of possible values of $D$ and $M$, respectively\footnote{Theoretically, $D$ has infinite possible values belonging to a contiguous space. But they are finite in a specific dataset. To simplify the notations, we use the discrete set $\mathcal{D}$ to represent the set of $D$'s possible values, and so is $M$.}. In particular, 
where $\mathcal{D}$ and $\mathcal{M}$ are the sample spaces of $D$ and $M$, respectively\footnote{Theoretically, $D$ has an infinite sample space. But the values are finite in a specific dataset. To simplify the notations, we use the discrete set $\mathcal{D}$ to represent the sample space of $D$, and so is $M$.}. In particular, 
Eq. (\ref{equ:P_Y_UI}a) follows the law of total probability; 
Eq. (\ref{equ:P_Y_UI}b) is obtained by Bayes rule;
% $M(\bm{d},\bm{u})$ in Eq. (\ref{equ:P_Y_UI}c) denotes the value of $M$ when $U=\bm{u}$ and $D=\bm{d}$. 
since $M$ can only take a value $M(\bm{d},\bm{u})$ if $U=\bm{u}$ and $D=\bm{d}$, \ie $P(M(\bm{d},\bm{u})|\bm{d},\bm{u})=1$, the sum over $\mathcal{M}$ in Eq. (\ref{equ:P_Y_UI}b) is removed;
$D$ is known if $U=\bm{u}$ is given. Thus the probability of $\bm{u}$ having the distribution $\bm{d}$ (\ie $P(\bm{d}|\bm{u})$) is $1$ if and only if $\bm{d}$ is $\bm{d}_u$; otherwise $P(\bm{d}|\bm{u}) = 0$, where $\bm{d}_u$ is the historical distribution of user $u$ over item groups. 

From Eq. (\ref{equ:P_Y_UI}d), we can find that $\bm{d}_u$ does not only affect the user representation $\bm{u}$ but also affects $Y$ via $M(\bm{d}_u,\bm{u})$, causing the spurious correlation: given the item $i$ in a group $g_n$, the more items in group $g_n$ the user $u$ has clicked in the history, the higher the prediction score $Y$ becomes. In other words, the high prediction scores are caused by the users' historical interest in the group instead of the items themselves. 
From the perspective of model prediction, $\bm{d}_u$ affects $\bm{u}$, which makes $\bm{u}$ favor the majority group. In Eq. (\ref{equ:P_Y_UI}d), a higher click frequency $p_u(g_n)$ in $\bm{d}_u$ will make $M(\bm{d}_u,\bm{u})$ represent a strong interest in group $g_n$, increasing the prediction scores of items in group $g_n$ via $P(Y|\bm{u},\bm{i},M(\bm{d}_u,\bm{u}))$. 
As such, the items in the majority group, even including the low-quality ones, are easy to have high prediction scores due to the effect of the confounder $D$. They occupy the recommendation opportunities of items in the minority group, and thus bias amplification happens.

% Indeed, the correlation inherits from the core idea of CF methods: recommending the items similar to the ones that users clicked in the history. Here, the similarity is that they belongs to the same item group.

\begin{figure}[tb]
\setlength{\abovecaptionskip}{0.1cm}
\setlength{\belowcaptionskip}{-0.40cm}
% \vspace{}
\centering
\includegraphics[scale=0.65]{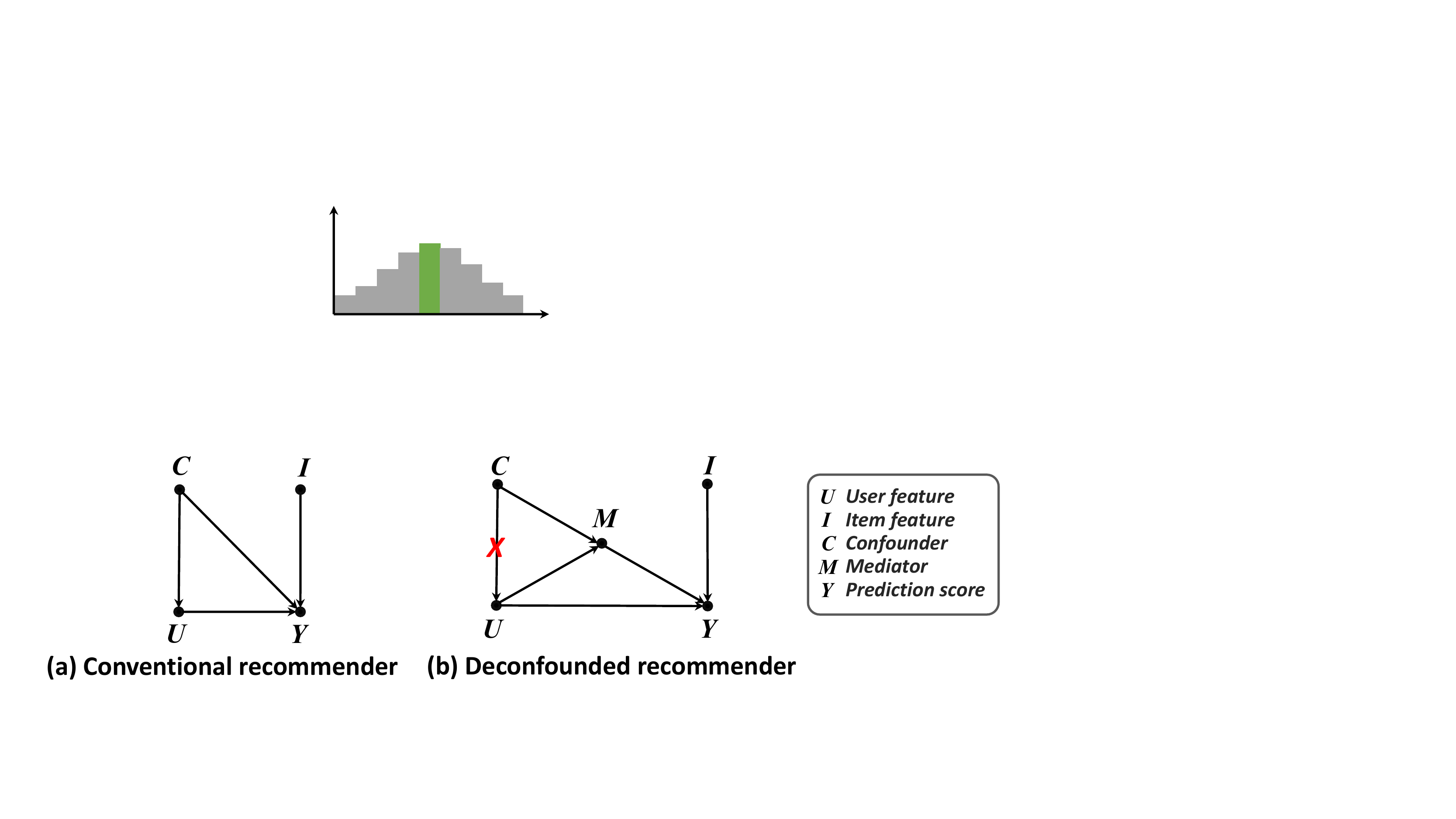}
\caption{(a) Illustration of user interest drift. (b) An example of the distribution of $D$ when the item group number is 2. Each node in the line represents a particular value $\bm{d}$, and a darker color denotes a higher probability of $\bm{d}$, \ie $P(\bm{d})$.}
\label{fig:interest_dirft}
\end{figure}

% Indeed, for some users, such correlation is beneficial to exclude the item groups they dislike, for example, users might only like action movies so that they seldom watched the movies in other minority groups in the history. 
The spurious correlation is harmful for most users because the items in the majority group are likely to dominate the recommendation list and narrow down the user interest. Besides, the undesirable and low-quality items in the majority group will dissatisfy users, leading to poor recommendation accuracy. 
Worse still, by analyzing Eq. \ref{equ:P_Y_UI}(d), we have a new observation: the prediction score $Y$ heavily relies on the user historical distribution over item groups, \ie $\bm{d}_u$. Once users' future interest in item groups changes (\ie user interest drift), the recommendations will be dissatisfying. For instance, as shown in Figure \ref{fig:interest_dirft}(a), the user interest in item groups is not stable, and thus the correlation caused by the confounder $D$ will not be reliable if the distribution $d_u$ is inconsistent between training and testing data.

\subsection{Deconfounded Recommender System}
To resolve the impact of the confounder, DecRS estimates the causal effect of user representation on the prediction score. Experimentally, the target can be achieved by collecting intervened data where the user representation is forcibly adjusted to eliminate the impact of confounder. However, such an experiment is too costly to achieve in large-scale and faces the risk of hurting user experience in practice. 
DecRS thus resorts to the causal technique: \textit{backdoor adjustment}~\cite{pearl2009causality, Pearl2018the, wang2018deconfounded}, which enables the estimation of causal effect from the observational data.
%backdoor adjustment with \textit{causal intervention} \cite{pearl2009causality, Pearl2018the} into the conventional CF framework. 
%To remove the spurious correlation caused by the confounder, DeCF introduces backdoor adjustment with \textit{causal intervention} \cite{pearl2009causality, Pearl2018the} into the conventional CF framework. 

% \vspace{-0.15cm}
% \paragraph{\textbf{Causal Intervention}} 
\subsubsection{Backdoor Adjustment}
%Different with the conventional RS that models the conditional probability $P(Y| U=\bm{u}, I=\bm{i})$, DecRS pushes the recommender models to estimate $P(Y| do(U=\bm{u}), I=\bm{i})$ by backdoor adjustment with causal intervention. 
According to the theory of backdoor adjustment~\cite{pearl2009causality}, the target of DecRS is formulated as: $P(Y| do(U=\bm{u}), I=\bm{i})$ where 
% $do(U=\bm{u})$ means forcefully setting the user representation as $\bm{u}$. Intuitively, 
$do(U=\bm{u})$ can be intuitively seen as cutting off the edge $D \rightarrow U$ in the causal graph and blocking the effect of $D$ on $U$ (\cf Figure \ref{fig:causal_graph}(b)).
%As shown in Figure \ref{fig:causal_graph}(b), causal intervention, formulated as $do(U=\bm{u})$, cuts off the edge $D \rightarrow U$ to remove the effect of $D$ on $U$ during training. 
% Thanks to the causal technique: \textit{backdoor adjustment}, we can estimate the causal-effect from results of intervention by observational data.
% As shown in Figure \ref{fig:causal_graph}(b), the recommender models are optimized to model $P(Y| do(U=\bm{u}), I=\bm{i})$. 
We then derive the specific expression of backdoor adjustment. Formally,

\begin{subequations}
\vspace{-0.4cm}
\label{equ:P_Y_do_UI}
\begin{align}\footnotesize
&P(Y|do(U=\bm{u}), I=\bm{i}) \notag \\
                &= \sum_{\bm{d}\in \mathcal{D}} P(\bm{d}|do(U=\bm{u}))P(Y|do(U=\bm{u}),\bm{i},M(\bm{d},do(U=\bm{u}))) \\
                &= \sum_{\bm{d}\in \mathcal{D}} P(\bm{d})P(Y|do(U=\bm{u}),\bm{i},M(\bm{d},do(U=\bm{u}))) \\
                &= \sum_{\bm{d}\in \mathcal{D}} P(\bm{d})P(Y|\bm{u},\bm{i},M(\bm{d},\bm{u})),
\end{align}
\end{subequations}
where the derivation of Eq. (\ref{equ:P_Y_do_UI}a) is the same as Eq. (\ref{equ:P_Y_UI}c), which follows the law of total probability and Bayes rule. Besides, Eq. (\ref{equ:P_Y_do_UI}b) and Eq. (\ref{equ:P_Y_do_UI}c) are obtained by two \textit{do} calculus rules:  \textit{insertion/deletion of actions} and \textit{action/observation exchange} in Theorem 3.4.1 of \cite{pearl2009causality}. 

%%%
% TODO: add back if it is essential.
%%%
%Thanks to \textit{do} calculus rules, we can estimate the results of intervention by observational data.

%From Figure \ref{fig:causal_graph}, we can find the main difference between the causal graphs used by conventional RS and DecRS is that the backdoor path $U \leftarrow D\rightarrow M \rightarrow Y$ is cut off in Figure \ref{fig:causal_graph}(b) so that 
%%%
% TODO: need to compare with equation 1d
%%%
As compared to Eq. \ref{equ:P_Y_UI}(d), DecRS estimates the prediction score with consideration of every possible value of $D$ subject to the prior $P(\bm{d})$, rather than the probability of $\bm{d}$ conditioned on $\bm{u}$.
%in Eq. (\ref{equ:P_Y_do_UI}c), the prediction score $Y$ doesn't rely on $\bm{d}_u$ in the training data. 
%In DecRS, it considers every possibility of $D$, subject to the prior $P(\bm{d})$. 
%%%
% TODO: replace click if wrong; replace instance if wrong.
%%%
Therefore, the items in the majority group will not receive high prediction scores purely because of a high click probability in $\bm{d}_u$. And thus backdoor adjustment alleviates bias amplification by removing the effect of $D$ on $U$.

%%%
% TODO: hard to understand and not very coherent, should focus on bias amplification. 
%%%
Intuitively, as shown in Figure \ref{fig:interest_dirft}(b), $D$ has extensive possible values in a specific dataset, \ie users have various historical distributions over item groups. 
% Each possible value can happen in a certain time period with a specific probability. 
In DecRS, the prediction score $Y$ considers various possible values of $D$. 
As such, 1) inevitably, DecRS removes the dependency on $\bm{d}_u$ in Eq. \ref{equ:P_Y_UI}(d)
and mitigates the spurious correlation,
% decreases the influence of $\bm{d}_u$ on the prediction score since the ``weight'' of $\bm{d}_u$ is reduced from $P(\bm{d}_u|\bm{u}) = 1$ to $P(\bm{d}_u) < 1$. 
and 2) theoretically, when user interest drift happens in the testing data, DecRS can produce a more robust and accurate prediction because the model has ``seen'' many different values of $D$ during training and doesn't heavily depend on the unreliable distribution $\bm{d}_u$ in Eq. \ref{equ:P_Y_UI}(d).

\subsubsection{Backdoor Adjustment Approximation}
%To perform backdoor adjustment, we need to explicitly model the causal relations between $U$, $I$, $D$, $M$, and $Y$. And then we estimate the prediction score via $P(Y| do(U=\bm{u}), I=\bm{i})$ in Eq. (\ref{equ:P_Y_do_UI}c). 
Theoretically, the sample space of $D$ is infinite, which makes the calculation of Eq. (\ref{equ:P_Y_do_UI}c) intractable. Therefore, it is essential to derive an efficient approximation of Eq. (\ref{equ:P_Y_do_UI}c).

$\bullet$ \textit{Sampling of $D$.} 
%To estimate $P(Y| do(U=\bm{u}), I=\bm{i})$ via Eq. (\ref{equ:P_Y_do_UI}c), we have to know each possible value $\bm{d}$ of $D$ and its probability $P(\bm{d})$. 
%%%
% this is wrong !
%%%
%However, the distribution of $D$ is inaccessible because theoretically $D$ has infinite possible values.
% users have various historical distributions over item groups during different periods. 
To estimate the distribution of $D$, we sample users' historical distributions over item groups in the training data, which comprise a discrete set $\mathcal{\Tilde{D}}$.
Formally, given a user $u$, $\bm{d}_u=[p_u(g_1), ..., p_u(g_N)] \in \mathcal{\Tilde{D}}$ and each click frequency $p_u(g_n)$ over group $g_n$ is calculated by
\begin{equation}
\label{equ:d_u_calc}
\begin{aligned}
p_u(g_n) &= \sum_{i\in \mathcal{I}} p(g_n|i)p(i|u) = \frac{\sum_{i\in \mathcal{H}_u}q_{g_n}^i}{|\mathcal{H}_u|},
\end{aligned}
\end{equation}
where $\mathcal{I}$ is the set of all items, $\mathcal{H}_u$ denotes the clicked item set by user $u$, and $q_{g_n}^i$ represents the probability of item $i$ belonging to group $g_n$. For instance, $\bm{q}^i = [1, 0, 0]$ with $q_{g_1}^i=1$ denotes that item $i$ only belongs to the first group. 
% We sample each user's historical distribution over item groups, comprising $\mathcal{\Tilde{D}}$, \ie a set of possible values of $D$ in the training data. 
In this work, we sample $D$ according to the user-item interactions in the training data, and thus the probability $P(\bm{d}_u)$ of user $u$ is obtained by $\frac{|\mathcal{H}_u|}{\sum_{v\in \mathcal{U}}\mathcal{H}_v}$ where $\mathcal{U}$ represents the user set. 
As such, we can estimate Eq. (\ref{equ:P_Y_do_UI}c) by 
\begin{equation}
\label{equ:E_D}
\begin{aligned}
P(Y|do(U=\bm{u}), I=\bm{i})  &\approx \sum_{\bm{d}\in \mathcal{\Tilde{D}}} P(\bm{d})P(Y|\bm{u},\bm{i},M(\bm{d},\bm{u}))\\
                    &= \sum_{\bm{d}\in \mathcal{\Tilde{D}}} P(\bm{d})f(\bm{u},\bm{i},M(\bm{d},\bm{u})),
\end{aligned}
\end{equation}
where each $\bm{d}$ is a distribution from one user, and we use a function $f(\cdot)$ (\eg FM~\cite{rendle2010factorization}) to calculate the conditional probability $P(Y|\bm{u},\bm{i},M(\bm{d},\bm{u}))$, similar to conventional recommender models.

$\bullet$ \textit{Approximation of \textit{ $\mathbb{E}_{\bm{d}}[f(\cdot)]$}.} 
The expected value of function $f(\cdot)$ of $\bm{d}$ in Eq. \ref{equ:E_D} is hard to compute because we need to calculate the results of $f(\cdot)$ for each $\bm{d}$ and the possible values in $\mathcal{\Tilde{D}}$ are extensive. A popular solution~\cite{abramovich2016some, wang2020visual} in statistics and machine learning theory is to make the approximation $\mathbb{E}_{\bm{d}}[f(\cdot)] \approx f(\bm{u},\bm{i},M(\mathbb{E}_{\bm{d}}[\bm{d}],\bm{u}))$.
% and show the error, \ie the \textit{Jensen gap} \cite{Gao2019bounds}, is small enough in the application scenario. 
Formally, the approximation takes the outer sum $\sum_{\bm{d}}P(\bm{d})f(\cdot)$ into the calculation within $f(\cdot)$:
\begin{equation}
\label{equ:E_appro}
\begin{aligned}
P(Y|do(U=\bm{u}), I=\bm{i})  &\approx f(\bm{u},\bm{i},M(\sum_{\bm{d}\in \mathcal{\tilde{D}}} P(\bm{d})\bm{d}, \bm{u})).\\
\end{aligned}
\end{equation}
The error of the approximation $\epsilon$ is measured by the \textit{Jensen gap}~\cite{abramovich2016some}:
\begin{equation}
\label{equ:E_error}
\begin{aligned}
\epsilon = |\mathbb{E}_{\bm{d}}[f(\cdot)] - f(\bm{u},\bm{i},M(\mathbb{E}_{\bm{d}}[\bm{d}],\bm{u}))|.\\
\end{aligned}
\end{equation}

%%%%%%%%%
% add error analysis
%%%%%%%%%

\begin{theorem}
\label{the:linear}
If $f$ is a linear function with a random variable $X$ as the input, then $E[f(X)] = f(E[X])$ holds under any probability distribution $P(X)$. Refer to \cite{Gao2019bounds, abramovich2016some} for the proof.
\end{theorem}

\begin{theorem}
\label{the:non_linear}
If a random variable $X$ with the probability distribution $P(X)$ has the expectation $\mu$, and the non-linear function $f:G\rightarrow \mathbb{R}$ where $G$ is a closed subset of $\mathbb{R}$, following:
\vspace{-0.1cm}
% \begin{itemize}[leftmargin=*]
\begin{itemize}
    \item [(1)] $f$ is bounded on any compact subset of $G$;
    \item [(2)] $|f(x)-f(\mu)|= O(|x-\mu|^{\beta})$ at $x\rightarrow \mu$ for $\beta>0$;
    \item [(3)] $|f(x)|= O(|x|^{\gamma})$ as $x\rightarrow +\infty$ for $\gamma\geq \beta$,
\end{itemize}
\vspace{-0.1cm}
then the inequality holds:
$|\mathbb{E}[f(X)] - f(\mu)| \leq T(\rho^{\beta}_{\beta} + \rho^{\gamma}_{\gamma}),$ 
where $\rho_{\beta} = \sqrt[\beta]{\mathbb{E}[|X-\mu|^{\beta}]}$, and $T=\text{sup}_{x\in G\backslash\{\mu\}}\frac{|f(x)-f(\mu)|}{|x-\mu|^{\beta}+|x-\mu|^{\gamma}}$ does not depend on $P(X)$. The proof can be found in \cite{Gao2019bounds}.
\end{theorem}

From Theorem \ref{the:linear}, we know that the error $\epsilon$ in Eq. \ref{equ:E_error} is zero if $f(\cdot)$ in Eq. \ref{equ:E_appro} is a linear function. However, most existing recommender models use non-linear functions to increase the representation capacity. In these cases, there is an upper bound of $\epsilon$ which can be estimated by Theorem \ref{the:non_linear}. It can be proven that the common non-linear functions in recommender models (\eg sigmoid in~\cite{rendle2010factorization}) satisfy the conditions in Theorem \ref{the:non_linear}, and the upper bound is small, especially when the distribution of $D$ concentrates around its expectation~\cite{Gao2019bounds}. 

\begin{table}[t]
% \vspace{-0.6cm}
\setlength{\abovecaptionskip}{0cm}
\setlength{\belowcaptionskip}{0.0cm}
\caption{Key notations and descriptions.}
\footnotesize
\label{tab:notation}
% \resizebox{0.5\textwidth}{!}{
\begin{tabular}{p{85pt}p{135pt}}
% \toprule
\specialrule{0.05em}{0pt}{1pt}
\textbf{Notation} & \textbf{Description} \\ \specialrule{0.05em}{1pt}{1pt}
% $U$ & User representation,\ie the embeddings of user features, \eg $\bm{u}$. \\ \hline
% $M$ & User interest representation over item groups.\\ \hline
% A specific value can be \\  $\textbf{m}_u=[\textbf{m}_{u, 1}, ..., \textbf{m}_{u, N}]$ where $\textbf{m}_{u, n}$ denotes the user interest in group $g_n$
$\bm{u}=[\bm{u}_1, ..., \bm{u}_K], \bm{u}_{k} \in \mathbb{R}^H$ & The representation vectors of $K$ user features. \\ \specialrule{0.0em}{1pt}{1pt}
$\bm{x}_u=[x_{u, 1}, ..., x_{u, K}]$ & 
\multicolumn{1}{m{140pt}}{The feature values of a user's $K$ features~\cite{rendle2010factorization}, \eg $[0.5, 1, ..., 0.2]$.} \\ \specialrule{0.0em}{1pt}{1pt}
$\bm{d}_u = [p_u(g_1), ..., p_u(g_N)]$ & 
\multicolumn{1}{m{140pt}}{$p_u(g_n)$ denotes the click frequency of user $u$ over group $g_n$ in the history, \eg $\bm{d}_u=[0.8, 0.2]$.} \\ \specialrule{0.0em}{1pt}{1pt}
% $\bm{m}_u = [\bm{m}_{u,1}, ..., \bm{m}_{u,N}]$ & 
% \multicolumn{1}{m{140pt}}{$\bm{m}_{u,n}$ represents the historical interest representation of user $u$ over group $g_n$.} \\ \specialrule{0.0em}{1pt}{1pt}
$\bm{m} = M(\bm{d}, \bm{u}) \in \mathbb{R}^H$ & 
\multicolumn{1}{m{140pt}}{The group-level representation of user $u$ under a historical distribution $\bm{d}$.} \\ \specialrule{0.0em}{1pt}{1pt}
$\mathcal{H}_u$ & The set of the items clicked by user $u$. \\ \specialrule{0.0em}{1pt}{1pt}
$\mathcal{U}, \mathcal{I}$ & The user and item sets, respectively. \\ \specialrule{0.0em}{1pt}{1pt}
$\bm{q}^i=[{q}^i_{g_1}, ..., {q}^i_{g_N}] \in \mathbb{R}^N$ & 
\multicolumn{1}{m{140pt}}{${q}^i_{g_n}$ denotes the probability of item $i$ belonging to group $g_n$, \eg $\bm{q}^i=[1, 0, 0]$.} \\ \specialrule{0.0em}{1pt}{1pt}
$\bm{v} = [\bm{v}_{1}, ..., \bm{v}_{N}], \bm{v}_{n} \in \mathbb{R}^H$ & 
\multicolumn{1}{m{140pt}}{$\bm{v}_{n}$ denotes the representation of group $g_n$.} \\ \specialrule{0.0em}{1pt}{1pt}
$\eta_u, \hat{\eta}_u$ & 
\multicolumn{1}{m{140pt}}{The symmetric KL divergence value of user $u$ and the normalized one, respectively.} \\ \specialrule{0.0em}{1pt}{1pt}
% \multicolumn{1}{m{140pt}}{$\bm{m}_{u,n}$ represents the historical interest representation of user $u$ over group $g_n$.} \\ \hline
\hline
%  \bottomrule
\end{tabular}
% }
\vspace{-0.3cm}
\end{table}

\subsection{Backdoor Adjustment Operator}
\label{sec:BA_operator}
% $\bullet$ \textit{Design of $f(\cdot)$.} 
To facilitate the usage of DecRS, we design the operator to instantiate backdoor adjustment, which can be easily plugged into recommender models to alleviate bias amplification. 
From Eq. \ref{equ:E_appro}, we can find that in addition to $\bm{u}$ and $\bm{i}$, $f(\cdot)$ takes $M(\bar{\bm{d}}, \bm{u})$ as the model input where $\bar{\bm{d}} = \sum_{\bm{d}\in \mathcal{\tilde{D}}} P(\bm{d})\bm{d}$. That is, if we can implement $M(\bar{\bm{d}}, \bm{u})$, existing recommender models can take it as one additional input to achieve backdoor adjustment.

% To estimate the prediction score via Eq. \ref{equ:E_appro}, we revise the existing recommender models $f(\cdot)$ which takes $M(\bar{\bm{d}}, \bm{u})$ where $\bar{\bm{d}} = \sum_{\bm{d}\in \mathcal{\tilde{D}}} P(\bm{d})\bm{d}$ as one additional input. Before that, we need to design the calculation of $M(\bar{\bm{d}}, \bm{u})$. 
% Recall that $M(\cdot)$ is the function to calculate the values of $M$ based on $D$ and $U$. 
Recall that $M$ denotes the group-level user representation which describes the user preference over item groups.
Given $\bar{\bm{d}}=[p(g_1), ..., p(g_N)]$, item group representation $\bm{v}=[\bm{v}_1, ..., \bm{v}_N]$, and user representation $\bm{u} = [\bm{u}_{1}, ..., \bm{u}_{K}]$ with feature values $\bm{x}_u = [{x}_{u,1}, ..., {x}_{u,K}]$~\cite{he2017nfm}, we calculate $M(\bar{\bm{d}}, \bm{u})$ by
\begin{equation}
\label{equ:M_ele_p}
\begin{aligned}
M(\bar{\bm{d}}, \bm{u}) = \sum_{a=1}^{N}\sum_{b=1}^{K} p(g_a)\bm{v}_a \odot x_{u,b} \bm{u}_{b},\\
\end{aligned}
\end{equation}
where $\odot$ denotes the element-wise product, and $\bm{v}_a \in \mathbb{R}^H$ is the item group representation for group $g_a$ proposed by us, which is randomly initialized like $\bm{u}$. The feature values in $\bm{x}_u$ are usually one, but in some special cases, it could be a float number. For instance, a user may have two jobs and the feature value for these two features can be set as 0.5 separately. 
Besides, we can also leverage a FM module~\cite{rendle2010factorization} or other high-order operators~\cite{feng2021cross}. 
Formally, we can obtain $\bm{w} = [\bar{\bm{d}}, \bm{x}_u] = [p(g_1), ..., p(g_N), {x}_{u,1}, ..., {x}_{u,K}]$ and $\bm{c} = [\bm{v}, \bm{u}] = [\bm{v}_1, ..., \bm{v}_N, \bm{u}_{1}, ..., \bm{u}_{K}]$ via concatenation, and then $M(\bar{\bm{d}}, \bm{u})$ can be calculated by a second-order FM module:
\begin{equation}
\label{equ:M_fm}
\begin{aligned}
M(\bar{\bm{d}}, \bm{u}) = \sum_{a=1}^{N+K}\sum_{b=1}^{N+K} w_a\bm{c}_a \odot w_{b} \bm{c}_{b},\\
\end{aligned}
\end{equation}
where $M(\bar{\bm{d}}, \bm{u})$ considers the interactions within $\bm{u}$ and $\bm{v}$ like FM, which is the main difference from Eq. \ref{equ:M_ele_p}. 
%%%
% Better to transfer a angle; take a passive gesture for DecRS. apply M to existing recommender models. Give an example.
%%%
Next, the group-level user representation $M(\bar{\bm{d}}, \bm{u})$ can be incorporated into existing recommender models as one additional user representation. Formally, if the generalized recommender models (\eg FM) are able to incorporate multiple feature representations, $M(\bar{\bm{d}}, \bm{u})$ is directly fed into the models to calculate $f(\bm{u},\bm{i}, M(\bar{\bm{d}},\bm{u}))$. Otherwise, $f(\cdot)$ can be implemented by a later-fusion manner, \ie $f(\cdot) = \delta * f^{'}(\bm{u},\bm{i}) + (1-\delta) * f^{'}(M(\bar{\bm{d}},\bm{u}),\bm{i})$ where $\delta$ is a hyperparameter and $f^{'}(\cdot)$ denotes the interaction module (\eg dot product) in recommender models to calculate the prediction score given user/item representations, such as neural collaborative filtering~\cite{He2017Neural}. 
Then the parameters $\theta$ in the recommender models are optimized by
\begin{equation}
\label{equ:loss}
\begin{aligned}
& \bar{\theta} =  \mathop{\arg\min}_{\theta}\sum_{(u, i,  \bar{y}_{u,i}) \in \bar{\mathcal{T}}}l(f(\bm{u},\bm{i},M(\bar{\bm{d}},\bm{u})), \bar{y}_{u,i}),
\end{aligned}
\end{equation}
where $\bar{y}_{u,i}\in \{0,1\}$ represents whether user $u$ has interacted with item $i$ (\ie $\bar{y}_{u,i}=1$) or not (\ie $\bar{y}_{u,i}=0$), $\mathcal{T}$ denotes the training data, and $l(\cdot)$ is the loss function, \eg log loss~\cite{He2017Neural}.
% As to the design of $f(\cdot)$, we can use any CF models that can take the representation $M(\bar{\bm{d}}, \bm{u})$ as the input, such as FM and Neural Factorization Machines (NFM).
%%%
% TODO: formula of training objective! Need to explain how to optimize the parameters !!!!
%%%

% \begin{equation}
% \begin{aligned}
% P(d_u) = \frac{|\mathcal{H}_u|}{\sum_{v\in \mathcal{U}}\mathcal{H}_v},
% \end{aligned}
% \end{equation}

% \subsection{In-depth Analysis}
% After the introduction of conventional training and deconfounded training, we give a comprehensive comparison and analysis of these two training manners.  

\subsection{Inference Strategy}
\label{sec:inference_strategy}

As mentioned before, DecRS alleviates bias amplification and produces more robust predictions when user interest drift happens. 
Indeed, for some users, bias amplification might be beneficial to exclude the item groups they dislike. For example, users might only like action movies so that they don't watch the movies in other groups. In these special cases, it makes sense to purely recommend extensive action movies. Therefore, it is better to develop a user-specific inference strategy to regulate the impact of backdoor adjustment dynamically. 
% As mentioned before, the correlation in conventional RS may help to exclude the irrelevant item groups for some users while backdoor adjustment in DecRS alleviates bias amplification and produces more robust predictions when user interest drift happens. 
% To combine their advantages, we develop an inference strategy to dynamically regulate the impact of backdoor adjustment during the inference stage.

% The user interest drift is unpredictable because it is affected by many factors, including the environment, social relations, and item popularity. 

% As discussed before, conventional training is more effective when the user interest is stable while deconfounded training can produce more robust recommendations if the temporal user interest drift happens frequently. 
By analyzing the user behavior, we find that many users have diverse interest and are likely to have interest drift while few users have stable interest in item groups over time (\eg only liking action movies). 
This inspires us to explore the user characteristics: is this user easy to change the interest distribution over item groups? Based on that, we propose a user-specific inference strategy for item ranking.
% Since the historical interactions are all we can have to observe user activities, we try to judge the user characteristics by the interest drift phenomenon in the history. 
If the user is easy to change the interest distribution over item groups in the history, we assume that he/she has diverse interest and will change it easily in future. And thus backdoor adjustment is essential to alleviate bias amplification. Otherwise, the impact of backdoor adjustment should be controlled.

% To study whether one user is likely to change interest, we analyze the change of the history distributions over time. 

% \paragraph{{Symmetric KL Divergence}}
$\bullet$ \textit{Symmetric KL Divergence.} 
% Since the historical interactions are all we can have to observe user activities, we try to judge the user characteristics by the interest drift phenomenon in the history. 
% assume that if the user's interest distribution over item groups changes dynamically in the history, the user will have a higher probability of interest drift in future. 
We employ the symmetric Kullback–Leibler (KL) divergence to quantify the user interest drift in the history. In detail, we divide the historical interaction sequence of user $u$ into two parts according to the timestamps. For each part, we calculate the historical distribution over item groups by Eq. \ref{equ:d_u_calc}, obtaining $\bm{d}_u^1=[p_u^1(g_1), ..., p_u^1(g_N)]$ and $\bm{d}_u^2=[p_u^2(g_1), ..., p_u^2(g_N)]$. Then, the distance between these two distributions is measured by the symmetric KL divergence: 
\begin{equation}
\label{equ:u_kl}
\begin{aligned}
\eta_u &= KL(\bm{d}_u^1|\bm{d}_u^2) +  KL(\bm{d}_u^2|\bm{d}_u^1) \\
       &= \sum_{n=1}^{N} P^1_u(g_n) \log\frac{P^1_u(g_n)}{P^2_u(g_n)} + \sum_{n=1}^{N} P^2_u(g_n) \log\frac{P^2_u(g_n)}{P^1_u(g_n)},
\end{aligned}
\end{equation}
where $\eta_u$ denotes the distribution distance of user $u$. A higher $\eta_u$ represents that the user is easier to change the interest distribution over item groups. Here, we only divide the historical interaction sequence into two parts to reduce the computation cost. More fine-grained division can be explored in future work if necessary.

% \subsubsection{{Auto-adjusting Inference}}

Based on the signal of $\eta_u$, 
% we can roughly judge whether backdoor adjustment should contribute more for the recommendation of user $u$. 
% As such, 
we utilize an inference strategy to adaptively fuse the prediction scores from the conventional RS and DecRS. Specifically, we first train the recommender model by $P(Y|U=\bm{u}, I=\bm{i})$ and $P(Y|do(U=\bm{u}), I=\bm{i})$, respectively, and their prediction scores are then automatically fused to regulate the impact of backdoor adjustment. Formally, 
% the score used to inference can be obtained by 
\begin{equation}
\label{equ:auto_infer}
\begin{aligned}
Y_{u,i} = (1-\hat{\eta}_u) * Y_{u,i}^{RS} + \hat{\eta}_u * Y_{u,i}^{DE},
\end{aligned}
\end{equation}
where $Y_{u,i}$ is the inference score for user $u$ and item $i$, $Y_{u,i}^{RS}$ and $Y_{u,i}^{DE}$ are the prediction scores from the conventional RS and DecRS, respectively. 
In particular, $\hat{\eta}_u$ is calculated by
\begin{equation}
\label{equ:auto_eta}
\begin{aligned}
\hat{\eta}_u = (\frac{\eta_u - \eta_{min}}{\eta_{max} - \eta_{min}})^\alpha
\end{aligned}
\end{equation}
where the normalized $\hat{\eta}_u \in [0,1]$, $\eta_{min}$ and $\eta_{max}$ are the minimum and maximum symmetric KL divergence values across all users, respectively. Besides, $\alpha \in [0, +\infty)$ is a hyper-parameter to further control the weights of $Y_{u,i}^{RS}$ and $Y_{u,i}^{DE}$ by human intervention. Specifically, $\hat{\eta}_u$ becomes larger if $\alpha \rightarrow 0$ due to $\hat{\eta}_u \in [0,1]$ which makes $Y_{u,i}$ favor $Y_{u,i}^{DE}$, and $\hat{\eta}_u$ decreases if $\alpha \rightarrow +\infty$.

% than $\frac{\eta_u - \eta_{min}}{\eta_{max} - \eta_{min}}$ if $\alpha < 1$ due to $\hat{\eta}_u \in [0,1]$, which will favor the predictions from DecRS. And $\hat{\eta}_u$ becomes smaller if $\alpha > 1$. 

From Eq. \ref{equ:auto_infer}, we can find that the inference for the users with high $\hat{\eta}_u$ will rely more on $Y_{u,i}^{DE}$. That is, $\eta_u$ automatically adjusts the balance between $Y_{u,i}^{RS}$ and $Y_{u,i}^{DE}$. 
% pursuing the history distribution over item groups and producing more robust recommendations to avoid the issue of user interest dirft.
Besides, we can regulate the impact of backdoor adjustment by tuning the hyper-parameter $\alpha$ in Eq. \ref{equ:auto_eta} for different datasets or recommender models. Theoretically, $\alpha$ is usually close to 0 because mitigating the spurious correlation improves the recommendation accuracy for most users.

To summarize, the proposed DecRS has three main differences from the conventional RS:
\begin{itemize}[leftmargin=*]
    \item DecRS models the causal effect 
    $P(Y | do(U=\bm{u}), I=\bm{i})$ instead of the conditional probability $P(Y | U=\bm{u}, I=\bm{i})$.
    \item DecRS equips the recommender models with a backdoor adjustment operator (\eg Equation \ref{equ:M_ele_p}).
    \item DecRS makes recommendations with a user-specific inference strategy instead of the simple model prediction (\eg a forward propagation).
\end{itemize}

\section{Related Work}

In this work, we explore how to alleviate bias amplification of recommender models by causal inference, which is highly related to fairness, diversity, and causal recommendation.

% \vspace{-0.15cm}
% \paragraph{Negative Effect of Bias Amplification}

% Existing CF recommenders are trained over the historical user feedback, and then deployed to serve users. Thereafter, the new user feedback is collected to re-train the recommenders, forming the feedback loop~\cite{}. 
% Due to the existence of feedback loop~\cite{chaney2018algorithmic, jiang2019degenerate}, bias amplification will become increasingly serious. The items in the majority group are recommended more and more frequently, leading to the Matthew Effect. Consequently, it will result in many negative issues: 1) narrowing down the user interest gradually, which is similar to the effect of \textit{filter bubbles}~\cite{pariser2011filter, bakshy2015exposure, nguyen2014exploring}. Worse still, the issue might evolve into \textit{echo chambers}~\cite{Ge2020Understanding, flaxman2016filter}, in which users' imbalanced interest are further reinforced by the repeated exposure to similar items; 2) low-quality items that users dislike might be recommended purely because it belongs to the majority group. They will occupy the recommendation opportunity of other high-quality items in the minority group, and 3) it is unfair for the items in the minority group which satisfy the users' expectations and deserve the exposure~\cite{singh2018fairness}. 
\vspace{0.1cm}
\noindent\textbf{Negative Effect of Bias Amplification.}
Due to the existence of feedback loop~\cite{chaney2018algorithmic}, bias amplification will become increasingly serious. 
Consequently, it will result in many negative issues: 1) narrowing down the user interest gradually, which is similar to the effect of \textit{filter bubbles}~\cite{nguyen2014exploring}. Worse still, the issue might evolve into \textit{echo chambers}~\cite{Ge2020Understanding}, in which users' imbalanced interest is further reinforced by the repeated exposure to similar items; 2) low-quality items that users dislike might be recommended purely because they are in the majority group, which deprive the recommendation opportunities of other high-quality items, causing low recommendation accuracy and unfairness.

% \vspace{-0.15cm}
% \paragraph{Fairness in Recommendation}
\vspace{0.1cm}
\noindent\textbf{Fairness in Recommendation.}
With the increasing attention on the fairness of machine learning algorithms~\cite{Kusner2017Counterfactual}, many works explore the definitions of fairness in recommendation and information retrieval~\cite{pitoura2020fairness, patro2020fairrec, mehrotra2018towards}. 
Generally speaking, they have two categories: individual fairness and group fairness. Individual fairness denotes that similar individuals (\eg users or items) should receive similar treatments (\eg exposure or clicks), such as amortized equity of attention~\cite{biega2018equity}. 
% Compared to individual fairness, group fairness is more widely studied because it is hard to quantify individual similarity and guarantee individual fairness in a single ranking~\cite{pitoura2020fairness, biega2018equity}. 
Besides, group fairness indicates that all groups are supposed to be treated fairly where individuals are divided into groups according to the protected attributes (\eg item category and user gender)~\cite{zhu2020measuring}. The particular definitions span from discounted cumulative fairness~\cite{yang2017measuring}, fairness of exposure~\cite{singh2018fairness}, to multi-sided fairness~\cite{burke2017multisided}. 
% When item features (\eg category) are used as protected attributes to divide items into groups, g
% In particular, group fairness might alleviate bias amplification in recommendation. 
% For example, fairness of exposure~\cite{singh2018fairness, Morik2020Controlling} pursues the fair exposure across different groups. 
% by which items in the majority group might be less recommended.

Another representative direction in fairness to reduce bias amplification is calibrated recommendation~\cite{steck2018calibrated}. It re-ranks the items to make the distribution of the recommended item groups follow the proportion in the browsing history. For example, if a user has watched 70\% action movies and 30\% romance movies, the recommendation list is expected to have the same proportion of movies. Although the fairness-related works, including calibrated recommendation, may alleviate bias amplification well, they are making the trade-off between ranking accuracy and fairness~\cite{steck2018calibrated, Morik2020Controlling, singh2018fairness}. The reason possibly lies in that they neglect the true cause of bias amplification.

% \vspace{-0.15cm}
% \paragraph{Diversity in Recommendation}
\vspace{0.1cm}
\noindent\textbf{Diversity in Recommendation.}
Diversity is regarded as one essential direction to get users out of filter bubbles in the information filtering systems~\cite{steck2018calibrated}. As to recommendation, diversity pursues the dissimilarity of the recommended items~\cite{Sun2020a, cheng2017learning}, where similarity can be measured by many factors, such as item category and embeddings~\cite{chandar2013preference}. However, most works might recommend many dissatisfying items when making diverse recommendations. 
% When the recommender model takes diversity as one ranking target, it can alleviate bias amplification because various items in the minority group will have a higher chance to be recommended. 
For example, the recommender model may trade off the accuracy to reduce the intra-list similarity by re-ranking~\cite{ziegler2005improving}.

% \vspace{-0.15cm}
% \paragraph{Causal Recommendation}
\vspace{0.1cm}
\noindent\textbf{Causal Recommendation.}
Causal inference has been widely used in many machine learning applications, spanning from computer vision~\cite{tang2020longtailed, niu2021counterfactual}, natural language processing~\cite{wu2020de, Feng2021Empowering, feng2020graph}, to information retrieval~\cite{bonner2018causal}. In recommendation, most works on causal inference~\cite{pearl2009causality} focus on debiasing various biases in user feedback, including position bias~\cite{Thorsten2017unbiased}, clickbait issue~\cite{wang2021click}, and popularity bias~\cite{zhang2021causal}. The most representative idea in the existing works is \textit{Inverse Propensity Scoring (IPS)}~\cite{Zhen2020Attribute, ai2018unbiased, wang2018deconfounded}, which first estimates the propensity score based on some assumptions, and then uses the inverse propensity score to re-weight the samples. For instance, Saito \etal estimated the exposure propensity for each user-item pair, and re-weighted the samples via IPS to solve the miss-not-at-random problem~\cite{saito2020unbiased}. However, IPS methods heavily rely on the accurate propensity estimation, and usually suffer from the high propensity variance. Thus it is often followed by the propensity clipping technique~\cite{saito2020unbiased, ai2018unbiased}. Another line of causal recommendation studies the effect of taking recommendations as treatments on user/system behaviors~\cite{zou2020counterfactual}, which is totally different from our work because we focus on the causal relations within the models. 

% mitigate dataset bias, remove the effect of confounders, 

\section{Experiments}
\label{sec:experiments}

We conduct extensive experiments to demonstrate the effectiveness of our DecRS by investigating the following research questions: 
\begin{itemize}[leftmargin=*]
    \item \textbf{RQ1:} How does the proposed DecRS perform across different users in terms of recommendation accuracy?
    \item \textbf{RQ2:} How does DecRS perform to alleviate bias amplification, compared to the state-of-the-art methods?
    \item \textbf{RQ3:} How do the different components affect the performance of DecRS, such as the inference strategy and the implementation of function $M(\cdot)$?
\end{itemize}

\begin{table}[t]
\setlength{\abovecaptionskip}{0cm}
\setlength{\belowcaptionskip}{00cm}
\caption{The statistics of the datasets.}
\label{tab:statistic_dataset}
\resizebox{.47\textwidth}{!}{
\begin{tabular}{l|l|l|l|l|l}
\hline
\textbf{Dataset} & \textbf{\#Users} & \textbf{\#Items} & \textbf{\#Interactions} & \textbf{\# Features} & \textbf{\#Group} \\ \hline
\textbf{ML-1M} & 3,883 & 6,040 & 575,276 & 13,408 & 18 \\ \hline
\textbf{Amazon-Book} & 29,115 & 16,845 & 1,712,409 & 46,213 & 253 \\ \hline
\end{tabular}
}
\vspace{-0.4cm}
\end{table}

\begin{table*}[t]
\setlength{\abovecaptionskip}{0cm}
\setlength{\belowcaptionskip}{0cm}
\caption{Overall performance comparison between DecRS and the baselines on ML-1M and Amazon-Book. \%improv. denotes the relative performance improvement achieved by DecRS over FM or NFM. The best results are highlighted in bold.}
\label{tab:overall_per}
\begin{center}
\setlength{\tabcolsep}{1mm}{
\resizebox{\textwidth}{!}{
\begin{tabular}{l|cccc|cccc|cccc|cccc}
\toprule
\textbf{} & \multicolumn{8}{c|}{\textbf{FM}} & \multicolumn{8}{c}{\textbf{NFM}} \\ \hline 
\multirow{2}{*}{\textbf{Method}} & \multicolumn{4}{c|}{\textbf{ML-1M}} & \multicolumn{4}{c|}{\textbf{Amazon-Book}} & \multicolumn{4}{c|}{\textbf{ML-1M}} & \multicolumn{4}{c}{\textbf{Amazon-Book}} \\
 & \textbf{R@10} & \textbf{R@20} & \textbf{N@10} & \textbf{N@20} & \textbf{R@10} & \textbf{R@20} & \textbf{N@10} & \textbf{N@20} & \textbf{R@10} & \textbf{R@20} & \textbf{N@10} & \textbf{N@20} & \textbf{R@10} & \textbf{R@20} & \textbf{N@10} & \textbf{N@20} \\ \hline 
\textbf{FM/NFM~\cite{rendle2010factorization, he2017nfm}} & 0.0676 & 0.1162 & 0.0566 & 0.0715 & 0.0213 & 0.0370 & 0.0134 & 0.0187 & 0.0659 & 0.1135 & 0.0551 & 0.0697 & 0.0222 & 0.0389 & 0.0144 & 0.0199 \\
\textbf{Unawareness~\cite{grgic2016case}} & 0.0679 & 0.1179 & 0.0575 & 0.0730 & 0.0216 & 0.0377 & 0.0138 & 0.0191 & 0.0648 & 0.1143 & 0.0556 & 0.0708 & 0.0206 & 0.0381 & 0.0133 & 0.0190 \\
\textbf{FairCo~\cite{Morik2020Controlling}} & 0.0676 & 0.1165 & 0.0570 & 0.0720 & 0.0212 & 0.0370 & 0.0135 & 0.0188 & 0.0651 & 0.1152 & 0.0554 & 0.0708 & 0.0219 & 0.0390 & 0.0142 & 0.0199 \\
\textbf{Calibration~\cite{steck2018calibrated}} & 0.0647 & 0.1149 & 0.0539 & 0.0695 & 0.0202 & 0.0359 & 0.0129 & 0.0181 & 0.0636 & 0.1131 & 0.0526 & 0.0682 & 0.0194 & 0.0335 & 0.0131 & 0.0178 \\ 
\textbf{Diversity~\cite{ziegler2005improving}} & 0.0670 & 0.1159 & 0.0555 & 0.0706 & 0.0207 & 0.0369 & 0.0131 & 0.0185 & 0.0641 & 0.1133 & 0.0540 & 0.0693 & 0.0215 & 0.0386 & 0.0140 & 0.0197 \\
\textbf{IPS~\cite{saito2020unbiased}} & 0.0663 & 0.1188 & 0.0556 & 0.0718 & 0.0213 & 0.0369 & 0.0135 & 0.0187 & 0.0648 & 0.1135 & 0.0544 & 0.0692 & 0.0213 & 0.0370 & 0.0137 & 0.0189 \\ \hline
\textbf{DecRS} & \textbf{0.0704} & \textbf{0.1231} & \textbf{0.0578} & \textbf{0.0737} & \textbf{0.0231} & \textbf{0.0405} & \textbf{0.0148} & \textbf{0.0205} & \textbf{0.0694} & \textbf{0.1218} & \textbf{0.0580} & \textbf{0.0742} & \textbf{0.0236} & \textbf{0.0413} & \textbf{0.0153} & \textbf{0.0211} \\
\textbf{\%improv.} & 4.14\% & 5.94\% & 2.12\% & 3.08\% & 8.45\% & 9.46\% & 10.45\% & 9.63\% & 5.31\% & 7.31\% & 5.26\% & 6.46\% & 6.31\% & 6.17\% & 6.25\% & 6.03\% \\
\bottomrule
\end{tabular}
}}
\end{center}
\vspace{-0.1cm}
\end{table*}

\begin{table*}[t]
\setlength{\abovecaptionskip}{0cm}
\setlength{\belowcaptionskip}{0cm}
\caption{Performance comparison across different user groups on ML-1M and Amazon-Book. Each line denotes the performance over the user group with $\eta_u >$ the threshold. We omit the results of threshold $> 4$ due to the similar trend.}
\label{tab:group_per}
\begin{center}
\resizebox{.9\textwidth}{!}{
\begin{tabular}{r|ccc|ccc|cccccc}
\toprule
\multicolumn{1}{l|}{\textbf{}} & \multicolumn{6}{c|}{\textbf{ML-1M}} & \multicolumn{6}{c}{\textbf{Amazon-Book}} \\ \hline
\textbf{FM} & \multicolumn{3}{c|}{\textbf{R@20}} & \multicolumn{3}{c|}{\textbf{N@20}} & \multicolumn{3}{c|}{\textbf{R@20}} & \multicolumn{3}{c}{\textbf{N@20}} \\
\textbf{Threshold} & \textbf{FM} & \textbf{DecRS} & \textbf{\%improv.} & \textbf{FM} & \textbf{DecRS} & \textbf{\%improv.} & \textbf{FM} & \textbf{DecRS} & \multicolumn{1}{c|}{\textbf{\%improv.}} & \textbf{FM} & \textbf{DecRS} & \textbf{\%improv.} \\ \hline
\textbf{0} & 0.1162 & 0.1231 & 5.94\% & 0.0715 & 0.0737 & 3.08\% & 0.0370 & 0.0405 & \multicolumn{1}{c|}{9.46\%} & 0.0187 & 0.0205 & 9.63\% \\
\textbf{0.5} & 0.1215 & 0.1296 & 6.67\% & 0.0704 & 0.0730 & 3.69\% & 0.0383 & 0.0424 & \multicolumn{1}{c|}{10.70\%} & 0.0192 & 0.0213 & 10.94\% \\
\textbf{1} & 0.1303 & 0.1412 & 8.37\% & 0.0707 & 0.0741 & 4.81\% & 0.0430 & 0.0479 & \multicolumn{1}{c|}{11.40\%} & 0.0208 & 0.0232 & 11.54\% \\
\textbf{2} & 0.1432 & 0.1646 & 14.94\% & 0.0706 & 0.0786 & 11.33\% & 0.0518 & 0.0595 & \multicolumn{1}{c|}{14.86\%} & 0.0231 & 0.0274 & 18.61\% \\
\textbf{3} & 0.1477 & 0.1637 & 10.83\% & 0.0620 & 0.0711 & 14.68\% & 0.0586 & 0.0684 & \multicolumn{1}{c|}{16.72\%} & 0.0256 & 0.0318 & 24.22\% \\
\textbf{4} & 0.1454 & 0.1768 & 21.60\% & 0.0595 & 0.0737 & 23.87\% & 0.0659 & 0.0793 & \multicolumn{1}{c|}{20.33\%} & 0.0284 & 0.0362 & 27.46\% \\ \hline
\textbf{NFM} & \multicolumn{3}{c|}{\textbf{R@20}} & \multicolumn{3}{c|}{\textbf{N@20}} & \multicolumn{3}{c|}{\textbf{R@20}} & \multicolumn{3}{c}{\textbf{N@20}} \\
\textbf{Threshold} & \textbf{NFM} & \textbf{DecRS} & \textbf{\%improv.} & \textbf{NFM} & \textbf{DecRS} & \textbf{\%improv.} & \textbf{NFM} & \textbf{DecRS} & \multicolumn{1}{c|}{\textbf{\%improv.}} & \textbf{NFM} & \textbf{DecRS} & \textbf{\%improv.} \\ \hline
\textbf{0} & 0.1135 & 0.1218 & 7.31\% & 0.0697 & 0.0742 & 6.46\% & 0.0389 & 0.0413 & \multicolumn{1}{c|}{6.17\%} & 0.0199 & 0.0211 & 6.03\% \\
\textbf{0.5} & 0.1187 & 0.1280 & 7.83\% & 0.0688 & 0.0735 & 6.83\% & 0.0401 & 0.0426 & \multicolumn{1}{c|}{6.23\%} & 0.0202 & 0.0218 & 7.92\% \\
\textbf{1} & 0.1272 & 0.1391 & 9.36\% & 0.0692 & 0.0747 & 7.95\% & 0.0438 & 0.0473 & \multicolumn{1}{c|}{7.99\%} & 0.0212 & 0.0234 & 10.38\% \\
\textbf{2} & 0.1452 & 0.1584 & 9.09\% & 0.0701 & 0.0771 & 9.99\% & 0.0530 & 0.0580 & \multicolumn{1}{c|}{9.43\%} & 0.0234 & 0.0269 & 14.96\% \\
\textbf{3} & 0.1478 & 0.1740 & 17.73\% & 0.0639 & 0.0723 & 13.15\% & 0.0614 & 0.0660 & \multicolumn{1}{c|}{7.49\%} & 0.0275 & 0.0319 & 16.00\% \\
\textbf{4} & 0.1442 & 0.1775 & 23.09\% & 0.0542 & 0.0699 & 28.97\% & 0.0709 & 0.0795 & \multicolumn{1}{c|}{12.13\%} & 0.0308 & 0.0371 & 20.45\% \\ 
\bottomrule
\end{tabular}
}
% }
\vspace{-0.25cm}
\end{center}
\end{table*}

\vspace{-0.1cm}
\subsection{Experimental Settings}

\vspace{2pt}
\noindent\textbf{Datasets.}
We use two benchmark datasets, ML-1M and Amazon-Book, in different real-world scenarios. 1) ML-1M is a movie recommendation dataset\footnote{\url{https://grouplens.org/datasets/movielens/1m/.}}, which involves rich user/item features, such as user gender, and movie genre.
We partition the items into groups according to the movie genre.
2) Amazon-Book is one of the Amazon product datasets\footnote{\url{https://jmcauley.ucsd.edu/data/amazon/.}}, where the book items can be divided into groups based on the book category (\eg sports).
To ensure data quality, we adopt the 20-core settings, \ie discarding the users and items with less than 20 interactions.
We summarize the statistics of datasets in Table \ref{tab:statistic_dataset}.

For each dataset, we sort the user-item interactions by the timestamps, and split them into the training, validation, and testing subsets with the ratio of $80\%$, $10\%$, and $10\%$.
For each interaction with the rating $\geq 4$, we treat it as a positive instance.
During training, we adopt the negative sampling strategy to randomly sample one item that the user did not interact with before as a negative instance.

% % \vspace{-0.15cm}
% \paragraph{Dataset}
% To evaluate the performance of DeCF, we perform experiments on two publicly available benchmark datasets: ML-1M and Amazon-Book. The statistics are in Table \ref{tab:statistic_dataset}. In particular, 1) ML-1M is a movie recommendation dataset\footnote{\url{https://grouplens.org/datasets/movielens/1m/.}}, including rich user/item features, \eg ID, user gender, and movie genre. The item groups are divided by the movie genre. 2) Amazon-Book is selected from the Amazon product datasets\footnote{\url{https://jmcauley.ucsd.edu/data/amazon/.}}, which has the book category to divide item groups, \eg sports books. We guarantee each user/item has at least 20 interactions to ensure the data quality~\cite{wang2019NGCF}.
% For two datasets, we treat the interactions with ratings $\geq 4$ as positive ones. Then we sort them by the timestamps, and split the interactions of each user into three groups by $80\%$, $10\%$, and $10\%$ as the training, validation, and testing data, respectively. For each positive interaction, we randomly sample one item the user never interacts with as the negative interaction during training. 

\vspace{5pt}
\noindent\textbf{Baselines.}
As our proposed DecRS is model-agnostic, we instantiate it on two representative recommender models, FM~\cite{rendle2010factorization} and NFM~\cite{he2017nfm}, to alleviate bias amplification and boost the predictive performance.
We compare DecRS with the state-of-the-art methods that might alleviate bias amplification of FM and NFM backbone models. In particular,
\begin{itemize}[leftmargin=*]
    \item \textbf{Unawareness}~\cite{grgic2016case, Kusner2017Counterfactual} removes the features of item groups (\eg movie genre in ML-1M) from the input of item representation $I$.
    \item \textbf{FairCo}~\cite{Morik2020Controlling} introduces one error term to control the exposure fairness across item groups. In this work, we calculate the error term based on the ranking list sorted by relevance, and its coefficient $\lambda$ in the ranking target is tuned in $\{0.01, 0.02, ..., 0.5\}$.
    \item \textbf{Calibration}~\cite{steck2018calibrated} is one state-of-the-art method to alleviate bias amplification. Specifically, it proposes a calibration metric $C_{KL}$ to measure the imbalance between the history and recommendation list, and minimizes $C_{KL}$ by re-ranking. Here the hyper-parameter $\lambda$ in the ranking target is searched in $\{0.01, 0.02, ..., 0.5\}$.
    \item \textbf{Diversity}~\cite{ziegler2005improving} aims to decrease the intra-list similarity, where the diversification factor is tuned in $\{0.01, 0.02, ..., 0.2\}$.
    \item \textbf{IPS}~\cite{saito2020unbiased} is a classical method in causal recommendation. Here we use $P(\bm{d}_u)$ as the propensity of user $u$ to down-weight the items in the majority group during debiasing training, and we employ the propensity clipping technique~\cite{saito2020unbiased} to reduce propensity variance, where the clipping threshold is searched in $\{2, 3, ..., 10\}$.
\end{itemize}

\vspace{-0.5pt}
\noindent\textbf{Evaluation Metrics.}
We evaluate the performance of all methods from two perspectives: recommendation accuracy and effectiveness of alleviating bias amplification.
In terms of accuracy, two widely-used metrics~\cite{wang2020disentangled}, Recall@K (R@K) and NDCG@K (N@K), are adopted under all ranking protocol~\cite{wang2019NGCF, wang2021denoising}, which test the top-K recommendations over all items that users never interact with in the training data.
As to alleviating bias amplification, we use the representative calibration metric $C_{KL}$ \cite{steck2018calibrated}, which quantifies the distribution drift over item groups between the history and the new recommendation list (comprised by the top-20 items).
Higher $C_{KL}$ scores suggest a more serious issue of bias amplification.

% \vspace{-0.15cm}
% \paragraph{Evaluation Metric}
% We evaluate the performance from two aspects: recommendation accuracy and the effectiveness of alleviating bias amplification. The former is evaluated by Recall@K (R@K) and NDCG@K (N@K) under all ranking protocol~\cite{wang2019NGCF}, \ie all items that users never interact with in the training data are candidates. The latter is evaluated by the calibration metric $C_{KL}$ proposed by \cite{steck2018calibrated}, which measures the distribution drift over item groups between the history and the new recommendation list (comprised by the top-20 items). Higher $C_{KL}$ scores represent that bias amplification is more serious. 

\vspace{3pt}
\noindent\textbf{Parameter Settings.}
We implement our DecRS in the PyTorch implementation of FM and NFM.
Closely following the original papers \cite{he2017nfm, rendle2010factorization}, we use the following settings: in FM and NFM, the embedding size of user/item features is 64, log loss~\cite{He2017Neural} is applied and the optimizer is set as Adagrad~\cite{duchi2011adaptive}; in NFM, a 64-dimension fully-connected layer is used. 
We adopt a grid search to tune their hyperparameters:
the learning rate is searched in $\{0.005, 0.01, 0.05\}$;
the batch size is tuned in $\{512, 1024, 2048\}$;
the normalization coefficient is searched in $\{0, 0.1, 0.2\}$,
and the dropout ratio is confirmed in $\{0.2, 0.3, ..., 0.5\}$.
Besides, $\alpha$ in the proposed inference strategy is tuned in $\{0.1, 0.2, ..., 10\}$, and the model performs the best in $\{0.2, 0.3, 0.4\}$, where $\alpha$ is close to 0, proving the advantages of our DecRS over the conventional RS as discussed in Section \ref{sec:inference_strategy}. 
We use Eq. \ref{equ:M_fm} to implement $M(\bar{\bm{d}}, \bm{u})$ and the backbone models take $M(\bar{\bm{d}}, \bm{u})$ as one additional feature. The exploration of the late-fusion manner is left to future work because it is not our main contribution.
% We utilize the proposed inference strategy to strengthen DecRS. 
Furthermore, we use the early stopping strategy~\cite{wei2019mmgcn, wang2018chat} --- stop training if R@10 on the validation set does not increase for 10 successive epochs.
For all approaches, we tune the hyper-parameters to choose the best models \wrt R@10 on the validation set, and report the results on the testing set. We released code and data at \url{https://github.com/WenjieWWJ/DecRS}.

% \vspace{-0.15cm}
% \paragraph{Hyper-parameter Settings}
% We implement DeCF based on the PyTorch implementation of FM and NFM. All the parameter initialization is the same with \cite{he2017nfm, rendle2010factorization}. The embedding size of user/item features is 64. And a 64-dimension fully-connected layer is used in NFM. For two models with different training methods, the learning rate and the batch size are tuned in $\{0.005, 0.01, 0.05\}$ and $\{512, 1024, 2048\}$, respectively. The normalization coefficient is searched in $\{0, 0.1, 0.2\}$ and dropout ratio is tuned in $\{0.2, 0.3, ..., 0.5\}$. The optimizer is set as Adagrad~\cite{duchi2011adaptive} following \cite{he2017nfm}. Besides, $\alpha$ in the auto-adjusting inference is searched in $\{0.1, 0.2, ..., 2\}$, and the model performs the best in $\{0.2, 0.3, 0.4\}$, \ie $\alpha < 1$, proving the advantages of deconfounded training. 
% Furthermore, we use early stopping to stop training if Recall@10 on the validation data does not increase for 10 successive epochs. For all the baselines and DeCF, we tune the same hyper-parameters to choose the best model \wrt Recall@10 on the validation data, and then evaluate it on the testing data by Recall, NDCG, and $C_{KL}$. 

\vspace{-0.25cm}
\subsection{Performance Comparison (RQ1 \& RQ2)}

\subsubsection{\textbf{Overall Performance \wrt Accuracy}}
We present the empirical results of all baselines and DecRS in Table \ref{tab:overall_per}.
Moreover, to further analyze the characteristics of DecRS, we split users into groups based on the symmetric KL divergence (\cf Eq. \ref{equ:u_kl}) and report the performance comparison over the user groups in Table \ref{tab:group_per}.
From the two tables, we have the following findings:
% The performance of the baselines and DeCF on two datasets is summarized in Table \ref{tab:overall_per}. In addition, to further analyze the characteristics of DeCF, we split users into groups based on the symmetric KL divergence calculated by Eq. \ref{equ:u_kl}. The performance comparison over user groups is shown in Table \ref{tab:group_per}.
% From two tables, we can observe that:
% \vspace{-0.2cm}
\begin{itemize}[leftmargin=*]
    \item Unawareness and FairCo only achieve comparable performance or marginal improvements over the vanilla FM and NFM on the two datasets. Possible reasons are the trade-offs among different user groups. To be more specific, for some users, discarding group features or preserving group fairness is able to reduce bias amplification and recommend more satisfying items. However, for most users with imbalanced interest in item groups, these approaches possibly recommend many disappointing items by pursuing group fairness.
    
    \item Calibration and Diversity perform worse than the vanilla backbone models, suggesting that simple re-ranking does hurt the recommendation accuracy. This is consistent with the findings in \cite{steck2018calibrated, ziegler2005improving}. Moreover, we ascribe the inferior performance of IPS to the inaccurate estimation and high variance of propensity scores. That is, the propensity cannot precisely estimate the effect of $D$ on $U$, even if the propensity clipping technique~\cite{saito2020unbiased} is applied.

    % \item The performance of Calibration, Diversity, and IPS based methods is inferior than the vanilla FM/NFM. It justifies that simply pursuing calibrated or diverse recommendations by re-ranking does hurt the recommendation accuracy, which is consistent with the findings in \cite{steck2018calibrated, ziegler2005improving}. We ascribe the inferior performance of IPS to the inaccurate estimation of propensity scores and the high propensity variance. Although we use the propensity clipping technique~\cite{saito2020unbiased}, the propensity cannot precisely estimate the effect of $D$ on $U$.
    
    \item DecRS effectively improves the recommendation performance of FM and NFM on the two datasets. As shown in Table \ref{tab:overall_per}, the relative improvements of DecRS over FM \wrt R@20 are $5.94\%$ and $9.46\%$ on ML-1M and Amazon-Book, respectively.
    This verifies the effectiveness of backdoor adjustment, which enables DecRS to remove the effect of confounder for many users. As a result, many less-interested or low-quality items from the majority group will not be recommended, thus increasing the accuracy.
    % thus reducing the bias amplification.
    
    % \item DeCF effectively improves the recommendation performance of FM and NFM on two datasets. For instance, in Table \ref{tab:overall_per}, the relative performance improvement of DeCF over FM on ML-1M and Amazon-Book in terms of Recall@20 is $5.94\%$ and $9.46\%$, respectively. This is because DeCF leverages deconfounded training to remove the effect of the confounder for many users. As such, many undesirable or low-quality items in the majority group will not be recommended. 
    
    \item As Table \ref{tab:group_per} shows, with the increase of $\eta_u$, the performance gap between DecRS and the backbone models becomes larger. For example, in the user group with $\eta_u>4$, the relative improvements \wrt N@20 over FM and NFM are $23.87\%$ and $28.97\%$, respectively.
    We attribute such improvements to the robust recommendation produced by DecRS.
    Specifically, DecRS equipped with backdoor adjustment is superior in reducing the spurious correlation and predicting users' diverse interest, especially for the users with the interest drift (\ie high $\eta_u$).
    % \item DeCF performs significantly better than FM/NFM over the users with high $\eta_u$. For example, the relative improvement over the user group with $\eta_u>4$ becomes larger than $20\%$ \wrt NDCG@20. It is attributed to the robust recommendations produced by DeCF. The users with high $\eta_u$ are easier to have the interest drift in the testing data, and DeCF with deconfounded training has the superiority to reduce the spurious correlation by causal intervention and predict users' diverse interest.  
    
\end{itemize}
% \vspace{-0.2cm}

\subsubsection{\textbf{Performance on Alleviating Bias Amplification}}
\begin{figure}[tb]
\setlength{\abovecaptionskip}{-0.10cm}
\setlength{\belowcaptionskip}{-0.4cm}
\centering 
\hspace{-0.3in}
    \subfigure{
    % \vspace{-0.1in}
    \label{subfig:a_kl_fm}
    \includegraphics[width=1.8in]{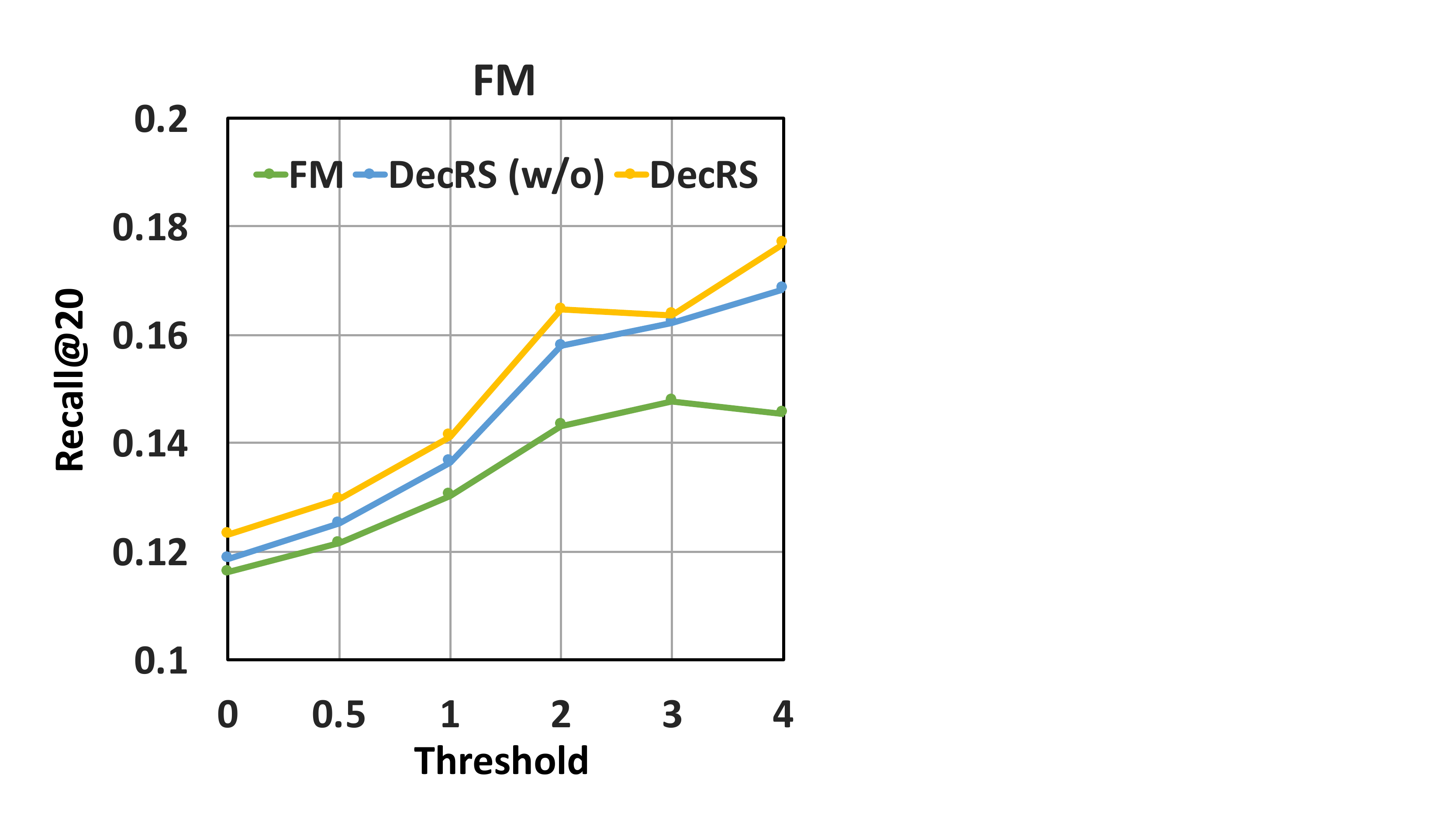}} 
    \hspace{-0.1in}
    \subfigure{
    % \vspace{-0.2in}
    \label{subfig:b_kl_nfm}
    \includegraphics[width=1.8in]{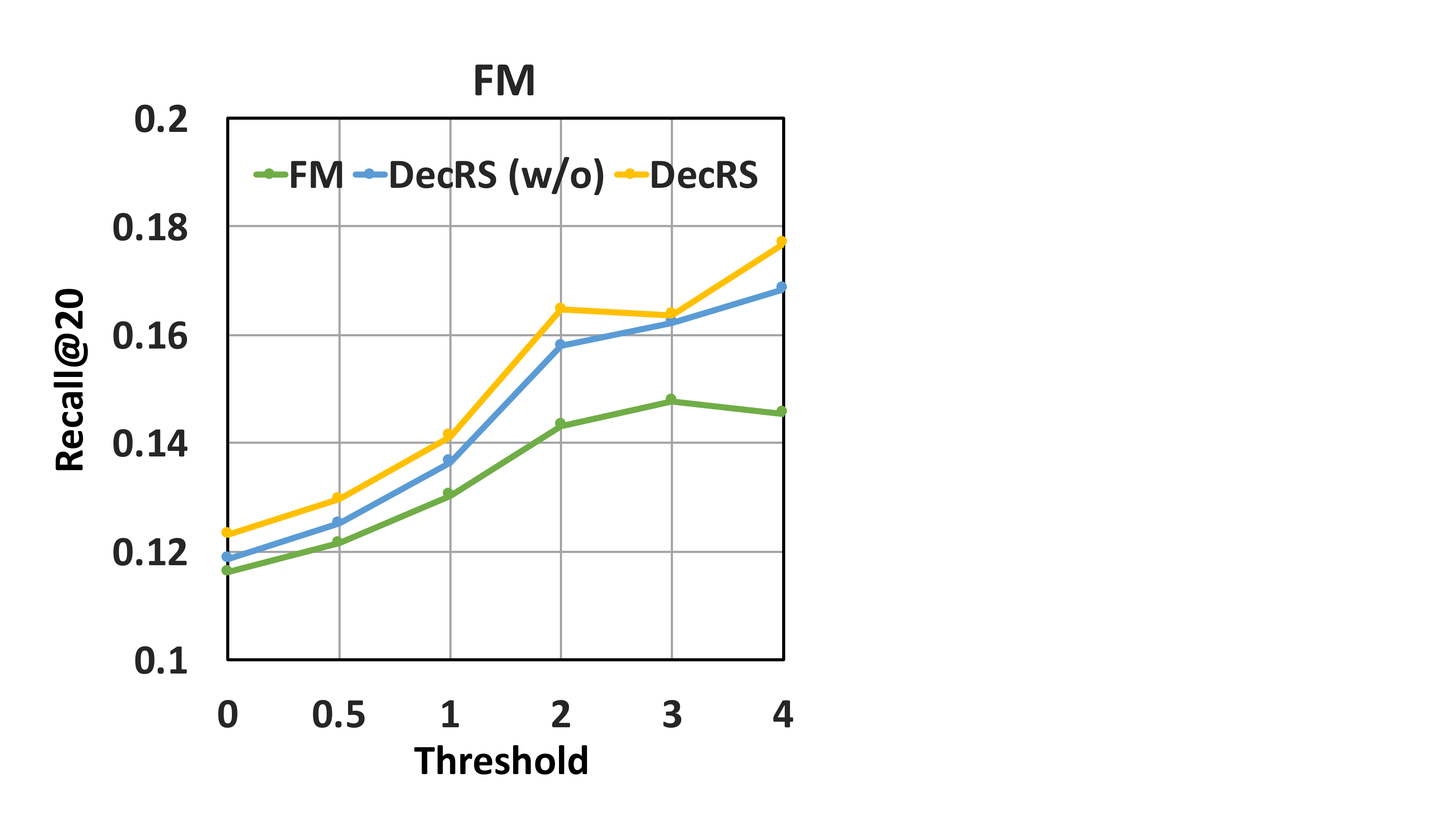}} 
    \hspace{-0.3in}
  \caption{The performance comparison between the baselines and DecRS on alleviating bias amplification.} 
    \label{fig:c_kl}
\end{figure}

In Figure \ref{fig:c_kl}, we present the performance comparison \wrt $C_{KL}$ between the vanilla FM/NFM, calibrated recommendation, and DecRS on ML-1M.
Due to space limitation, we omit other baselines that perform worse than calibrated recommendation and the results on Amazon-Book which have similar trends.
We have the following observations from Figure \ref{fig:c_kl}.
1) As compared to the vanilla models, calibrated recommendation achieves lower $C_{KL}$ scores, suggesting that the bias amplification is reduced.
However, it comes at the cost of lower recommendation accuracy, as shown in Table \ref{tab:overall_per}.
2) Our DecRS consistently achieves lower $C_{KL}$ scores than calibrated recommendation across all user groups.
More importantly, DecRS does not hurt the recommendation accuracy.
This evidently shows that DecRS solves the bias amplification problem well by 
embracing causal modeling for recommendation, and justifies the effectiveness of backdoor adjustment on reducing spurious correlations. 

% The performance comparison \wrt $C_{KL}$ between the vanilla FM/NFM, calibrated recommendation, and DeCF on ML-1M is presented in Figure \ref{fig:c_kl}. The results of other baselines are not shown because they perform worse than the state-of-the-art calibrated recommendation. Besides, the performance on Amazon-Book with the similar trend is omitted to save space. From Figure \ref{fig:c_kl}, we find that 1) calibrated recommendation alleviates bias amplification of the CF models. Considering the results in Table \ref{tab:overall_per}, we know it achieves this at the expense of sacrificing accuracy; and 2) the proposed DeCF framework performs better than calibrated recommendation on alleviating bias amplification across different user groups. More importantly, DeCF doesn't hurt the recommendation accuracy because it solves the problem by fundamentally changing the training method, which justifies the effectiveness of deconfounded training on reducing spurious correlation. 

\subsection{In-depth Analysis (RQ3)}

\subsubsection{\textbf{Effect of the Inference Strategy}}
\begin{figure}[tb]
\setlength{\abovecaptionskip}{-0.10cm}
\setlength{\belowcaptionskip}{0.0cm}
\centering 
\hspace{-0.3in}
    \subfigure{
    % \vspace{-0.1in}
    \label{subfig:a_r10_fm}
    \includegraphics[width=1.8in]{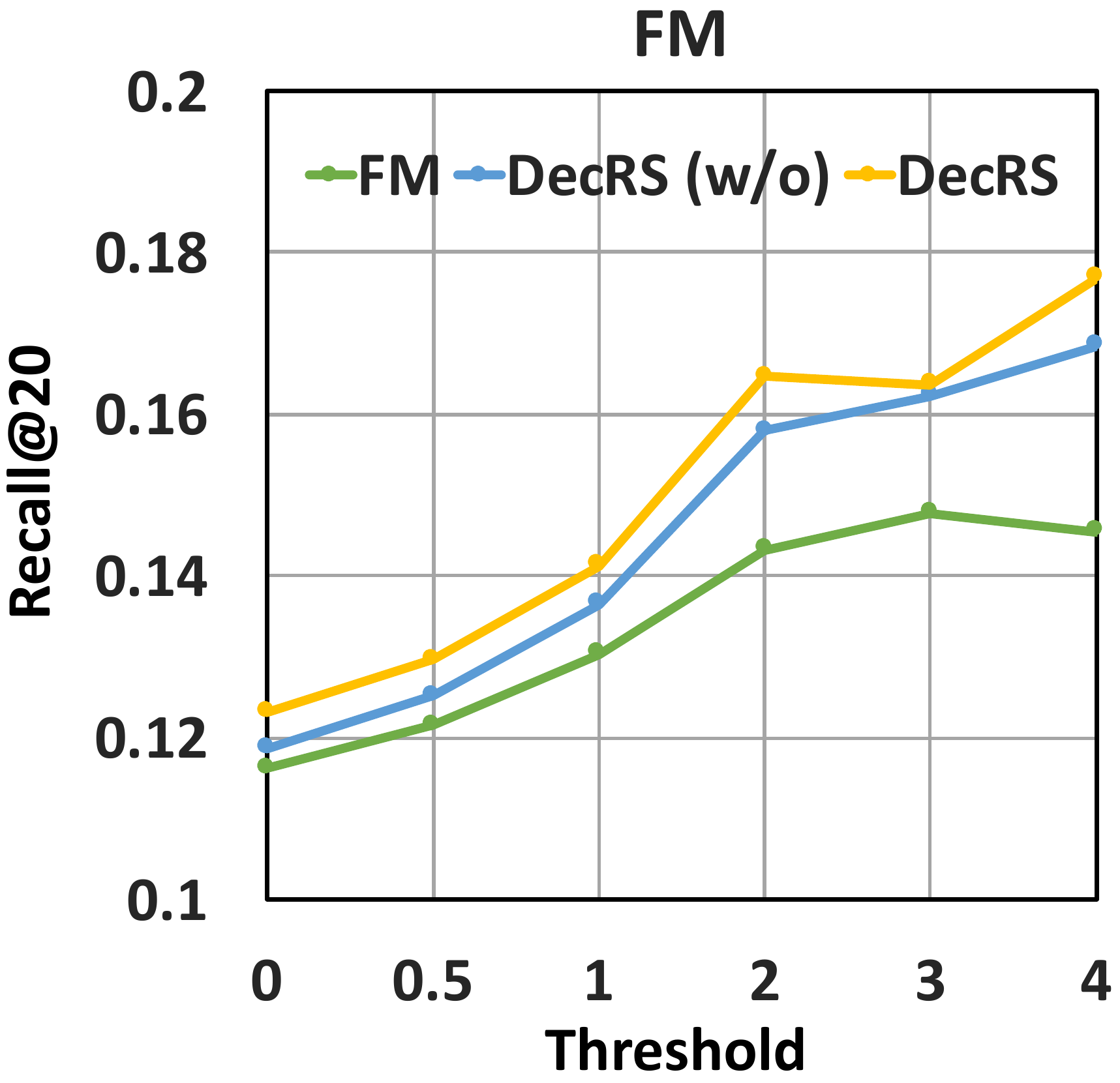}} 
    \hspace{-0.1in}
    \subfigure{
    % \vspace{-0.2in}
    \label{subfig:b_r10_nfm}
    \includegraphics[width=1.8in]{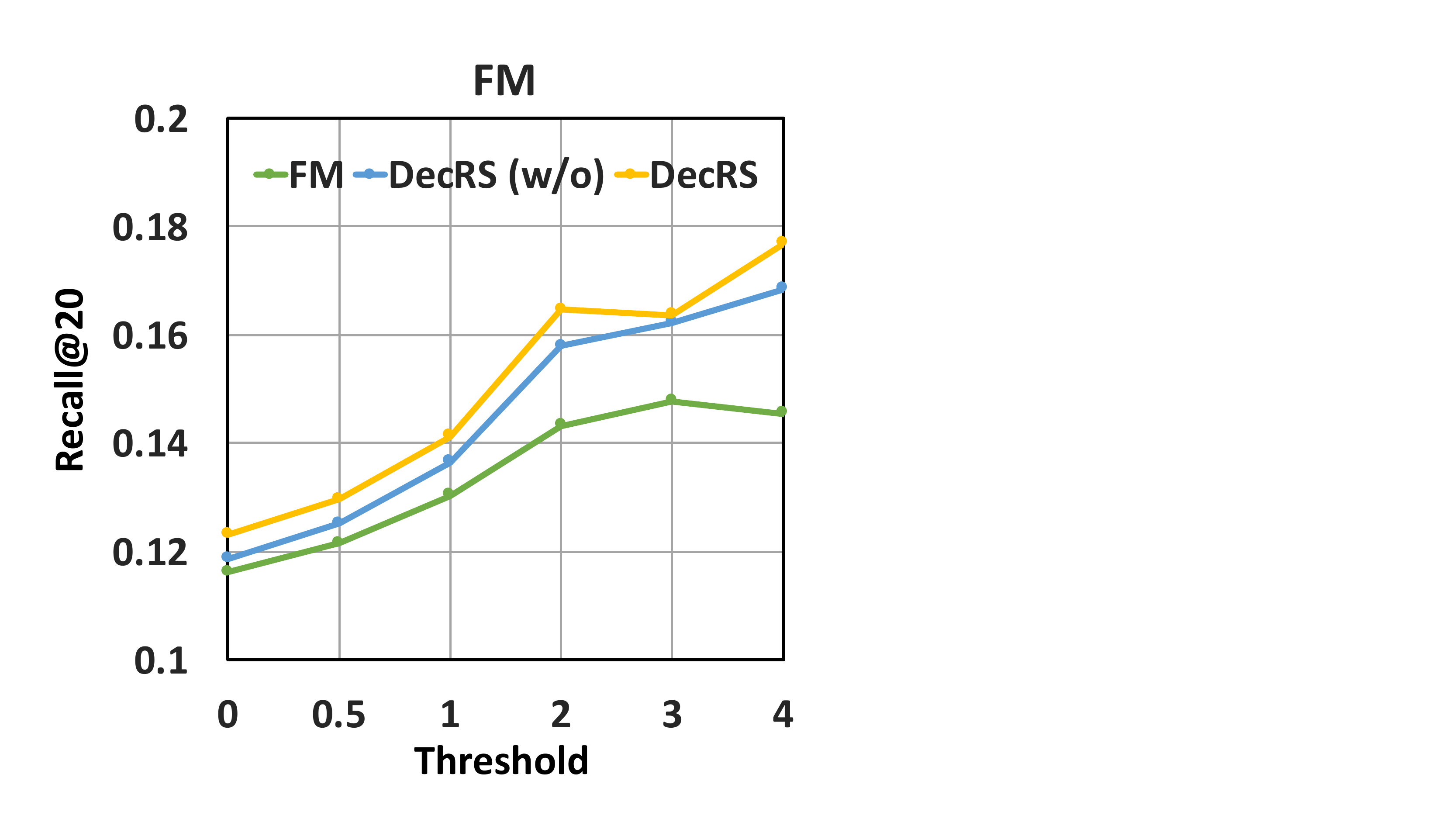}} 
    \hspace{-0.3in}
  \caption{Ablation study of DecRS on ML-1M. 
%   ``DecRS (w/o)'' denotes the results of disabling the inference strategy.
  } 
    \label{fig:r10_dt}
\end{figure}

We first answer the question: Is it of importance to conduct the inference strategy for DecRS?
Towards this end, one variant ``DecRS (w/o)'' is constructed by disabling the inference strategy and only using the prediction $Y^{DE}$ in Eq. \ref{equ:auto_infer} for inference.
We illustrate its results in Figure \ref{fig:r10_dt} with the following key findings.
1) The performance of ``DecRS (w/o)'' drops as compared with that of DecRS, indicating the effectiveness of the inference strategy.
% It also validates the rationality of leveraging the correlation in conventional RS to exclude the irrelevant item groups for some users.
2) ``DecRS (w/o)'' still outperforms FM and NFM consistently, especially over the users with high $\eta_u$. This suggests the superiority of DecRS over the conventional RS.
It achieves more accurate predictions of user interest by mitigating the effect of the confounder via backdoor adjustment approximation. 

% To prove the utility of auto-adjusting inference, we remove it from the DeCF framework, and only utilize the prediction $Y^{DT}$ from deconfounded training for inference. The results are labeled with ``DT'' and visualized in Figure \ref{fig:r10_dt}. From the figure, we have the following observations: 1) the performance of DT drops as compared with that of DeCF, proving the effectiveness of auto-adjusting inference. It validates the rationality of leveraging the advantages of conventional training to exclude the irrelevant item groups. 2) DT still outperforms FM/NFM, especially over the users with high $\eta_u$, which shows the superiority of deconfounded training over conventional training. It achieves more accurate predictions of user interests by mitigating the effect of the confounder. 

\begin{table}[t]
\vspace{-0.1cm}
\setlength{\abovecaptionskip}{0cm}
\setlength{\belowcaptionskip}{-0.2cm}
\caption{Effect of the design of $M(\cdot)$.}
\label{tab:effect_m}
\resizebox{0.35\textwidth}{!}{
\begin{tabular}{l|l|l|l|l}
\hline
\textbf{Method} & \textbf{R@10} & \textbf{R@20} & \textbf{N@10} & \textbf{N@20} \\ \hline
\textbf{FM} & 0.0676 & 0.1162 & 0.0566 & 0.0715 \\ \hline
\textbf{DecRS-EP} & 0.0685 & 0.1205 & 0.0573 & 0.0730 \\ \hline
\textbf{DecRS-FM} & 0.0704 & 0.1231 & 0.0578 & 0.0737 \\ \hline
\end{tabular}
% }
}
\vspace{-0.4cm}
\end{table}

\subsubsection{\textbf{Effect of the Implementation of $M(\cdot)$}}
As mentioned in Section \ref{sec:BA_operator}, we can implement the function $M(\cdot)$ by either Eq. \ref{equ:M_ele_p} or Eq. \ref{equ:M_fm}.
We investigate the influence of different implementations and construct two variants, DecRS-EP and DecRS-FM, which employ the element-wise product in Eq. \ref{equ:M_ele_p} and the FM module in Eq. \ref{equ:M_fm}, respectively.
We summarize their performance comparison over FM on ML-1M in Table \ref{tab:effect_m}. 
While being inferior to DecRS-FM, DecRS-EP still performs better than FM. 
This proves the superiority of DecRS-FM over DecRS-EP, and also shows that DecRS with different implementations still surpasses the vanilla backbone models, which further suggests the stability and effectiveness of DecRS.

\section{Conclusion and Future Work}
\label{sec:discussion}
In this work, we explained that bias amplification in recommender models is caused by the confounder from a causal view. To alleviate bias amplification, we proposed a novel DecRS with an approximation operator for backdoor adjustment. 
DecRS explicitly models the causal relations in recommender models, and leverages backdoor adjustment to remove the spurious correlation caused by the confounder. Besides, we developed an inference strategy to regulate the impact of backdoor adjustment. Extensive experiments validate the effectiveness of DecRS on alleviating bias amplification and improving recommendation accuracy. 

This work takes the first step to incorporate backdoor adjustment into existing recommender models. In future, there are many research directions that deserve our attention. 1) The discovery of more fine-grained causal relations in recommendation models. This work starts to mitigate the spurious correlations caused by the confounder while recommendation is an extremely complex scenario, involving many observed/hidden variables that are waiting for causal discovery. 2) The proposed DecRS has the potential to reduce various biases in information retrieval and recommendation, such as position bias and popularity bias. The causes of the biases are also related to the imbalanced training data. 3) Bias amplification is one essential cause of the filter bubble~\cite{nguyen2014exploring} and echo chambers~\cite{Ge2020Understanding}. The effect of DecRS on mitigating these issues can be studied in future work.

{
\tiny
\bibliographystyle{ACM-Reference-Format}
\balance
\bibliography{bibfile}
}

\end{document}